\documentclass[aps, amsmath, amssymb, superscriptaddress, nofootinbib, reprint, preprintnumbers]{revtex4-1}
\usepackage{graphicx}
\usepackage{epsfig}
\usepackage{epstopdf}
\usepackage{dcolumn}
\usepackage{bm}
\usepackage{subfigure}
\usepackage[dvipsnames]{xcolor} 
\usepackage{booktabs}
\usepackage{txfonts} 
\usepackage{lineno} 
\usepackage{bbold} 
\usepackage{amsmath}
\usepackage[fleqn]{mathtools}
\usepackage{hyperref}
\usepackage{slashed} 

\usepackage{tikz,tikz-feynman}
\tikzfeynmanset{compat=1.1.0, warn luatex=false}
\usetikzlibrary{arrows, arrows.meta}

\usepackage{placeins}

\newcommand{\psibar}{\ensuremath{\overline\psi} }

\newcommand{\Fpgg}{\mathcal{F}_{P\gamma\gamma}}
\newcommand{\Fpigg}{\mathcal{F}_{\pi^0\gamma\gamma}}

\newcommand{\amupi}{a_\mu^{\pi^0-\mathrm{pole}}}

\newcommand{\chibar}{\ensuremath{\overline\chi}}

\newcommand{\tsep}{t_{\mathrm{sep}}}

\newcommand{\Fig}{FIG.}
\newcommand{\Tab}{Tab.}
\newcommand{\Tabs}{Tabs.}
\newcommand{\Eq}{Eq.}
\newcommand{\Eqs}{Eqs.}

\newcommand{\app}[1]{Appendix \ref{#1}}

\graphicspath{{fig/}}

\begin{document}

\title{Pion Transition Form Factor from Twisted-Mass Lattice QCD \texorpdfstring{\\}{} and the Hadronic Light-by-Light \texorpdfstring{$\pi^0$}{pion}-pole Contribution to the Muon \texorpdfstring{$g-2$}{g-2}}

\author{C.~Alexandrou}
\affiliation{Department of Physics, University of Cyprus, 20537~Nicosia, Cyprus}
\affiliation{Computation-based Science and Technology Research Center, The Cyprus Institute, 20~Konstantinou Kavafi Street, 2121~Nicosia, Cyprus}
   
\author{S.~Bacchio}
\affiliation{Computation-based Science and Technology Research Center, The Cyprus Institute, 20~Konstantinou Kavafi Street, 2121~Nicosia, Cyprus}

\author{G.~Bergner}
\affiliation{University of Jena, Institute for Theoretical Physics, Max-Wien-Platz 1, D-07743~Jena, Germany}

\author{S.~Burri}
\affiliation{Albert Einstein Center for Fundamental Physics,
  Institute for Theoretical Physics, University of Bern,
  Sidlerstrasse 5,
  CH--3012 Bern, Switzerland}

\author{J.~Finkenrath}
\affiliation{Bergische Universit{\"a}t Wuppertal, Gau{\ss}straße 20, 42119 Wuppertal, Germany}
\affiliation{Computation-based Science and Technology Research Center, The Cyprus Institute, 20~Konstantinou Kavafi Street, 2121~Nicosia, Cyprus}

\author{A.~Gasbarro}
\affiliation{Albert Einstein Center for Fundamental Physics,
  Institute for Theoretical Physics, University of Bern,
  Sidlerstrasse 5,
  CH--3012 Bern, Switzerland}
  
\author{K.~Hadjiyiannakou}
\affiliation{Department of Physics, University of Cyprus, 20537~Nicosia, Cyprus}
\affiliation{Computation-based Science and Technology Research Center, The Cyprus Institute, 20~Konstantinou Kavafi Street, 2121~Nicosia, Cyprus}

\author{K.~Jansen}
\affiliation{Deutsches Elektronen-Synchrotron DESY, Platanenallee
  6, 15738 Zeuthen, Germany}
  
\author{G.~Kanwar}
\affiliation{Albert Einstein Center for Fundamental Physics,
  Institute for Theoretical Physics, University of Bern,
  Sidlerstrasse 5,
  CH--3012 Bern, Switzerland}
  
\author{B.~Kostrzewa}
\affiliation{High Performance Computing and Analytics Lab,
  Rheinische Friedrich-Wilhelms-Universit{\"a}t Bonn,
  Friedrich-Hirzebruch-Allee~8,
  D-53115 Bonn, Germany}
  
\author{G.~Koutsou}
\affiliation{Computation-based Science and Technology Research Center, The Cyprus Institute, 20~Konstantinou Kavafi Street, 2121~Nicosia, Cyprus}

\author{K.~Ottnad}
\affiliation{PRISMA$^+$ Cluster of Excellence and Institut f{\"u}r Kernphysik,
Johannes Gutenberg-Universit{\"a}t Mainz, Johann-Joachim-Becher-Weg
45, D-55128 Mainz, Germany}

\author{M.~Petschlies}
\affiliation{HISKP (Theory), Rheinische
  Friedrich-Wilhelms-Universit{\"a}t Bonn,
  Nussallee~14-16,
  D-53115~Bonn, Germany}
\affiliation{Bethe Center for Theoretical Physics, Rheinische
  Friedrich-Wilhelms-Universit{\"a}t Bonn, Wegelerstraße 10, D-53115 Bonn, Germany}
  
\author{F.~Pittler}
\affiliation{Computation-based Science and Technology Research Center, The Cyprus Institute, 20~Konstantinou Kavafi Street, 2121~Nicosia, Cyprus}

\author{F.~Steffens}
\affiliation{HISKP (Theory), Rheinische
  Friedrich-Wilhelms-Universit{\"a}t Bonn,
  Nussallee~14-16,
  D-53115~Bonn, Germany}
\affiliation{Bethe Center for Theoretical Physics, Rheinische
  Friedrich-Wilhelms-Universit{\"a}t Bonn, Wegelerstraße 10, D-53115 Bonn, Germany}
  
\author{C.~Urbach}
\affiliation{HISKP (Theory), Rheinische
  Friedrich-Wilhelms-Universit{\"a}t Bonn,
  Nussallee~14-16,
  D-53115~Bonn, Germany}
\affiliation{Bethe Center for Theoretical Physics, Rheinische
  Friedrich-Wilhelms-Universit{\"a}t Bonn, Wegelerstraße 10, D-53115 Bonn, Germany}

\author{U.~Wenger}
\affiliation{Albert Einstein Center for Fundamental Physics,
  Institute for Theoretical Physics, University of Bern,
  Sidlerstrasse 5,
  CH--3012 Bern, Switzerland}

\collaboration{Extended Twisted Mass Collaboration}

\date{\today}

\begin{abstract}
The neutral pion generates the leading pole contribution to the hadronic light-by-light tensor, which is given in terms of the nonperturbative transition form factor $\Fpigg(q_1^2,q_2^2)$.  
Here we present an ab-initio lattice calculation of this quantity in the continuum and at the physical point using twisted-mass lattice QCD. 
We report our results for the transition form factor parameterized using a model-independent conformal expansion valid for arbitrary space-like kinematics and compare it with experimental measurements of the single-virtual form factor, the two-photon decay width, and the slope parameter. 
We then use the transition form factors to compute the pion-pole contribution to the hadronic light-by-light scattering in the muon $g-2$, finding $\amupi = 56.7(3.2) \times 10^{-11}$.
\end{abstract}

\maketitle

\section{\label{sec:intro}Introduction}
The anomalous magnetic moment of the muon (the muon $g-2$) provides a
stringent test of the Standard Model at high precision and the
possibility to glimpse subtle effects of potential new physics beyond the Standard Model.
Recent results from the Fermilab experiment
\cite{PhysRevLett.126.141801,Muong-2:2023cdq} in combination with the
Brookhaven result \cite{Bennett:2006fi} have yielded the most precise
experimental determination to date at the level of 19 ppm: $a_\mu(\text{exp}) = 116\,592\,059(22) \times 10^{-11}$, where $a_\mu \equiv (g - 2)_\mu / 2$.
A comparable precision is striven for on the theoretical side with the
recent state of affairs
summarized in the white paper \cite{Aoyama:2020ynm}.  

The leading hadronic contributions to the muon anomalous magnetic
moment come from diagrams involving the hadronic vacuum polarization
(HVP) and the hadronic light-by-light (HLbL) scattering tensors.
Both make significant contributions at the level of precision achieved
by the experimental results. It is important to improve the
theoretical determinations of both contributions to match future
targets of experimental precision; this work addresses the HLbL contribution. 

Two complementary approaches to the theoretical determination of the HLbL contribution are becoming recognizable for reliable estimates. On one hand, 
several direct lattice calculations have analyzed the complete HLbL tensor \cite{Blum:2014oka,Blum:2015gfa,Blum:2016lnc,Blum:2017cer,Chao2021,Chao:2022xzg,Asmussen2022}.
On the other hand, the HLbL tensor can be decomposed into contributions from exchanges of various intermediate on-shell states 
in a data-driven dispersive approach, where contributions from heavier intermediate states are suppressed \cite{Colangelo2014a,Colangelo:2014pva,Colangelo:2015ama,Pauk:2014rfa}. 
The latter approach has provided the most precise theoretical estimates of the HLbL contribution to the muon $g-2$ to date~\cite{Aoyama:2020ynm},
while the former has the benefit of allowing for a fully
self-contained lattice QCD calculation from first principles
without resorting to chiral effective field theory and experimental data.
However, there has been an ongoing effort to provide also ab-initio data based on lattice QCD calculations
for the leading contributions within the dispersive decomposition of the HLbL.

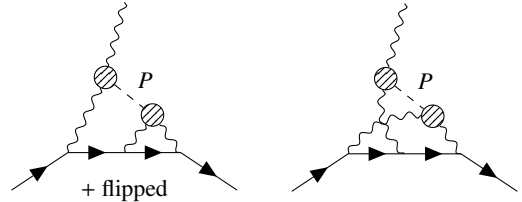
\begin{figure}
    \centering
    \tikzfeynmanset{
  tff/.style={
    /tikz/shape=circle,
    /tikz/draw=black,
    /tikz/pattern=north east lines
  }
}

\tikzfeynmanset{
  blob/.style={
    /tikz/shape=circle,
    /tikz/minimum size=1cm,
    /tikz/draw=black,
    /tikz/pattern=north east lines
  }
}

%
%
\begin{tikzpicture}[baseline=(current bounding box.center), scale=1.0]
\begin{feynman}[medium]

\vertex (i) at (0.5,0.5);
\vertex (a) at (1.25,1);
\vertex (b) at (2,1);
\vertex (c) at (2.75,1);
\vertex (o) at (3.5,0.5);
\vertex [tff] (x) at (1.75,2) {};
\vertex [tff] (z) at (2.375,1.5) {};
\vertex (y) at (2,3);

\diagram* {
(i) -- [fermion] (a) -- [fermion] (b) -- [fermion] (c) -- [fermion] (o),
(a) -- [boson] (x) -- [boson] (y),
(b) -- [boson] (z) -- [boson] (c),
(x) -- [scalar, edge label=\(P\)] (z)
};
\end{feynman}
\node at (2,0.5) {+ flipped};
\end{tikzpicture}
\qquad
\begin{tikzpicture}[baseline=(current bounding box.center), scale=1.0]
\begin{feynman}[medium]

\vertex (i) at (0.5,0.5);
\vertex (a) at (1.25,1);
\vertex (b) at (2,1);
\vertex (c) at (2.75,1);
\vertex (o) at (3.5,0.5);
\vertex [tff] (x) at (1.75,2) {};
\vertex [tff] (z) at (2.375,1.5) {};
\vertex (y) at (2,3);

\diagram* {
(i) -- [fermion] (a) -- [fermion] (b) -- [fermion] (c) -- [fermion] (o),
(b) -- [boson, bend left] (x) -- [boson] (y),
(a) -- [boson, bend left] (z) -- [boson] (c),
(x) -- [scalar, edge label=\(P\)] (z)
};
\end{feynman}
\node at (2,0.5) {\phantom{+ crossed}};
\end{tikzpicture}
    \caption{Pseudoscalar-pole diagrams contributing to the leading
      order hadronic light-by-light (HLbL )scattering. Each striped circle represents a nonperturbative $P \rightarrow \gamma^* \gamma^*$ transition form factor required to evaluate these contributions, where ``$P$'' indicates the exchanged pseudoscalar meson. The ``flipped'' contribution signifies the diagram with the pseudoscalar and photon portions of the diagram horizontally flipped.
    }
    \label{fig:HLbL_pole_dominance}
\end{figure}

The leading single-pole contributions to the HLbL component of the
muon $g-2$ arise from the exchange of neutral pseudoscalar mesons $P \in \{\pi^0, \eta, \eta'\}$ 
as shown in \Fig~\ref{fig:HLbL_pole_dominance}. The key nonperturbative input to these diagrams are the transition form factors (TFFs) $\Fpgg(q_1^2, q_2^2)$, defined 
by \cite{Knecht:2001qf}
\begin{equation}
\begin{aligned}
&\epsilon_{\mu\nu\alpha\beta}q^\alpha p^\beta \Fpgg(q_1^2,q_2^2) \equiv M_{\mu \nu}(p, q_1) \\
&\hspace{40pt} = i \int d^4 x e^{i q_1 x} \langle 0 | T\{j_\mu(x) j_\nu(0) \}|P(p)\rangle ,
\label{eq:minkowskiFF}
\end{aligned}
\end{equation}
where
\begin{equation}
    j_\mu(x) = \psibar(x) \gamma_\mu Q \psi(x)
\end{equation}
is the electromagnetic current written in terms of the charge matrix
$Q$. The momentum of the pseudoscalar meson satisfies the on-shell condition $p^2 = M_P^2$, while the photon momenta $q_1$ and $q_2 = p-q_1$ may be either on-shell or virtual.
The pole contribution to the HLbL tensor from the
pseudoscalar meson $P$ is given by
\begin{equation}
\begin{aligned}
\Pi^{(P)}_{\mu\nu\lambda\rho}(q_1,q_2,q_3) &= i \frac{\Fpgg(q_1^2,q_2^2)\Fpgg(q_3^2,(q_1+q_2+q_3)^2)}{(q_1+q_2)^2-M_P^2}\\
&\quad \times \epsilon_{\mu\nu\alpha\beta} q_1^\alpha q_2^\beta \epsilon_{\lambda\rho \sigma \tau} q_3^\sigma (q_1+q_2)^\tau \\
&\quad + (q_1 \leftrightarrow q_3) + (q_2 \leftrightarrow q_3).
\end{aligned}
\end{equation}
Integrating over space-like $q_1$, $q_2$, and $q_3 = -q_1 - q_2$ with the appropriate kinematical factors from the diagrams in \Fig~\ref{fig:HLbL_pole_dominance} yields contributions to the total muon $g-2$ from each pseudoscalar meson exchange.

The $\pi^0$-pole contribution to the muon $g-2$ is larger than those of the $\eta$ and $\eta'$ by roughly a factor of four~\cite{PhysRevD.94.053006,Jegerlehner2009}. This feature, together with the relative simplicity of accessing the pion TFF in both lattice and data-driven approaches, means that the study of this transition form factor is farthest advanced at this point.
Direct experimental data for $\Fpigg(q_1^2, q_2^2)$ is primarily available
in the singly virtual (either $q_1^2 = 0$ or $q_2^2 = 0$) regime with large photon virtuality~\cite{CELLO:1990klc,CLEO:1997fho,BaBar:2009rrj, BaBar:2011nrp,Belle:Uehara2012}. Fits to this data using Canterbury approximants have yielded a data-based estimate across all virtualities~\cite{Masjuan2017a}.
Meanwhile, the pion TFF has also been determined 
from indirect experimental data using a dispersive approach \cite{Hoferichter:2018dmo,Hoferichter:2018kwz}.
On the other hand, an ab-initio determination is desirable to remove
the dependence on experimental inputs. As such, the neutral
pseudoscalar meson
TFF has also been computed on the lattice. The calculation of
  the neutral-pion TFF on the lattice has been pioneered by the
Mainz group using Wilson quarks
\cite{Gerardin:2016cqj,Gerardin:2019vio}. The BMW collaboration
recently added their results using staggered quarks \cite{Gerardin:2023naa}. Our preliminary
results with Wilson twisted-mass quarks appeared in Refs.~\cite{Burri:2021cxr,BurriPoS2022}.
The major sources of uncertainty in these approaches
is the extrapolation to the continuum and, in case the quark masses are
not tuned to their physical values as in Refs.~\cite{Gerardin:2016cqj,Gerardin:2019vio}, the extrapolation to the physical point.  A first
  calculation of the $\eta$-meson TFF and the corresponding pole
  contribution at one lattice spacing has been reported by our collaboration in
  Ref.~\cite{Alexandrou:2022qyf} and results for both the $\eta$- and $\eta'$-meson TFF
  and pole contributions addressing all systematic errors have been
  presented by the BMW
  collaboration in Ref.~\cite{Gerardin:2023naa}.

In the present work, we complement these existing efforts by
evaluating the pion transition form factor in the continuum using the twisted-mass
lattice QCD discretization, with ensembles generated directly at the physical point. This calculation serves to strengthen the lattice consensus of the continuum limit and minimizes systematic effects from extrapolation of the pion mass.
In this work, we present a parameterization of the pion transition form factor and use it to calculate the $\pi^0$-pole contribution to the muon $g-2$. As our main result, we find $\amupi = 56.7(3.2) \times 10^{-11}$. 
We further use the transition form factor to estimate the two-photon decay width as $\Gamma(\pi \to \gamma \gamma) = 7.50(0.50)\,\mathrm{eV}$, and the slope parameter as $b_\pi = 2.16(0.20)\,\mathrm{GeV}^{-2}$, 
which provides input for determining the electromagnetic interaction radius. Quoted uncertainties include statistical errors and systematic uncertainty from analysis choices and the continuum extrapolation.

The structure of this paper is as follows. 
In Sec.~\ref{sec:methodology}, we review the theoretical background for the extraction of the required amplitude from lattice QCD simulations. 
In Sec.~\ref{sec:TFF}, we describe the calculation of the transition form factors and their extrapolation to photon virtualities required for the observables that we calculate. 
In Sec.~\ref{sec:observables}, we present the calculation of $\amupi$
from the transition form factors and the determination of the decay width $\Gamma(\pi \to \gamma \gamma)$ and the slope parameter $b_\pi$. 
Next, in Sec.~\ref{sec:model_averaging}, we discuss the estimation
of statistical and systematic errors based on model averaging,
before presenting our results in Sec.~\ref{sec:results}, comparing them to experimental and theoretical predictions. 
Then, in Sec.~\ref{sec:conclusion}, we make some concluding remarks and give an outlook on possible future directions for improving the calculation presented here. 

\section{Euclidean amplitude}
\label{sec:methodology}
By analytic continuation, the Minkowski neutral pion transition form factor in \Eq~\eqref{eq:minkowskiFF} is related to the Euclidean amplitude
\begin{equation}
\label{eq:tilde_A_timeordering}
\tilde{A}_{\mu \nu}(\tau) = \left<0 \big| T\{j_\mu(\vec{q}_1, \tau) j_\nu(\vec{p}-\vec{q}_1, 0) \} \big| \pi^0(p)\right>
\end{equation}
by~\cite{Ji:2001wha,Ji:2001nf,Dudek:2006ut}
\begin{equation}
\label{eq:matelem_eucl}
M_{\mu \nu}(p, q_1) = i^{n_0} M_{\mu \nu}^E(p, q_1), \quad M_{\mu \nu}^E = \int_{-\infty}^{\infty} d\tau \, e^{\omega_1 \tau} \tilde{A}_{\mu \nu}(\tau),
\end{equation}
where $n_0 = \delta_{\mu,0}+\delta_{\nu,0}$, $q_1 = (\omega_1,
\vec{q}_1)$, and $\tau$ indicates a Euclidean time coordinate. Note
that $\tilde{A}_{\mu\nu}$ depends on the momenta $\vec{q}_1$ and
$\vec{p}$, but this dependence is suppressed for conciseness of notation.
The analytic continuation is valid for virtualities $q^2_1$ and $q^2_2$ below the two-pion threshold at $4 m_\pi^2$~\cite{Ji:2001wha}. In particular, this includes the entire space-like quadrant $q_{1,2}^2 \leq 0$ required for evaluation of the muon $g-2$.

\subsection{Three-point function}
We access the amplitude $\tilde{A}_{\mu\nu}(\tau)$ through the following lattice three-point correlation function
\begin{equation}
  \begin{aligned}
    C_{\mu \nu} (\tau, t_{\pi}) &= a^6 \sum_{\vec{x}, \vec{z}} \left\langle j_\mu(\vec{x},\tau) \; j_\nu(\vec{0},0) \; P^{0\dag}(\vec{z},-t_{\pi}) \right\rangle e^{-i \vec{q}_1 \cdot \vec{x}} e^{i \vec{p} \cdot \vec{z}} \\
    &= \left\langle j_\mu(\vec{q}_1, \tau) \; j_\nu(\vec{0}, 0) \; P^{0\dag}(\vec{p}, -t_{\pi}) \right\rangle,
    \label{eq:amplitude}
  \end{aligned}
\end{equation}
where $P^{0\dag}$ is an interpolating operator for the neutral pion and $\langle \dots \rangle$ indicates the Euclidean path-integral expectation value.
Translation invariance, or equivalently conservation of momentum, ensures that the current $j_\nu$ must carry momentum $\vec{q}_2 = \vec{p} - \vec{q}_1$.
We denote the minimal separation between the pion creation operator and the nearest electromagnetic current by $\tsep = \mathrm{min}(\tau,0) + t_{\pi}$.
When $\tsep \gg 0$, or equivalently $t_{\pi} \gg 0,-\tau$, the pion creation operator $P^{0\dag}(\vec{p}, -t_{\pi})$ is in the distant Euclidean past, and the Euclidean matrix element in \Eq~\eqref{eq:tilde_A_timeordering} can be recovered from the three-point function as
\begin{equation}
\begin{aligned}
	C_{\mu\nu}(\tau,t_{\pi}) &\stackrel{\tsep \gg 0}{=}
	e^{-E_\pi t_{\pi}} \frac{Z_\pi}{2E_\pi} \left\langle 0 \left| T\{ j_\mu(\tau) j_\nu(0) \} \right| \pi^0(\vec{p}) \right\rangle \\
    &\qquad + O(e^{-\Delta E \tsep}) \\
    &= e^{-E_\pi t_\pi} \frac{Z_\pi}{2E_\pi} \tilde{A}_{\mu\nu}(\tau) + O(e^{-\Delta E \tsep}),
\end{aligned}
\end{equation}
where
\begin{equation}
Z_\pi \equiv \left< 0 | P^0(\vec{0}, 0) | \pi^0(\vec{p}) \right>
\end{equation}
and $\Delta E$ indicates the energy gap to the first excited state produced by $P^{0\dag}(\vec{p},-t_{\pi})$.
Given this limiting behavior, we define 
\begin{equation} \label{eq:A_tilde}
  \begin{aligned}
        \tilde{A}_{\mu\nu}(\tau;t_\pi) &\equiv \frac{2 E_\pi}{Z_\pi} e^{E_\pi t_\pi} C_{\mu\nu}(\tau,t_\pi) \\
        &= \tilde{A}_{\mu\nu}(\tau) + O(e^{-\Delta E \tsep}).
      \end{aligned}
\end{equation}
In practice, we compute this quantity for finite values of $t_{\pi}$
taken sufficiently large to ensure the excited state contamination is
suppressed below our uncertainties. When it is clear from context, 
we elide the $t_\pi$ argument in the following.

The parameters $E_\pi$ and $Z_\pi$ are extracted from a separate
analysis of the the $\tau$-dependence of the two-point function
\begin{equation}
\begin{aligned}
    C^{(2)}(\tau) &\equiv a^3 \sum_{\vec{x}} e^{-i \vec{p} \cdot \vec{x}} \left< 0 | P^0(\vec{x},\tau) P^{0\dag}(\vec{0},0) | 0 \right> \\
    &= \frac{|Z_\pi|^2}{2 E_\pi} e^{-E_\pi \tau} + O(e^{-\Delta E \tau}).
  \end{aligned}
  \label{eq:pion_2-pt-function}
\end{equation}
We note that for maximally twisted fermions the value of $Z_\pi$
is strictly positive and can be unambiguously extracted from
\Eq~\eqref{eq:pion_2-pt-function}, because it is  related to the pion decay constant $f_\pi$ through $Z_\pi = f_\pi m_\pi
\sinh(m_\pi)/2 \mu_l$ where $\mu_l$ is the twisted light-quark mass
parameter \cite{Frezzotti:2003ni,Frezzotti:2000nk}.

\subsection{Twisted-mass Lattice QCD}
We perform our calculation using twisted-basis fermion fields, denoted $\chi$. For the light quarks, these are related to the physical-basis doublet $\psi = (u,d)$ through the maximal twist transformation
\begin{equation}
  \psi = e^{i\pi\gamma_5\tau^3/4} \, \chi \,,\,\,\,
\bar{\psi}= \bar{\chi} \, e^{i\pi\gamma_5\tau^3/4} \,,
\end{equation}
where $\tau^3$ is the generator of the third component of isospin, represented by the third Pauli matrix.
In the continuum limit, the twist transformation is a trivial change of basis. At finite lattice spacing, however, the Wilson term breaks chiral invariance meaning the twisted basis provides an alternative path towards the continuum limit, with certain desirable properties such as automatic $O(a)$ improvement \cite{Frezzotti:2003ni, Frezzotti:2004wz}.

To relate calculations performed in the twisted basis to the physical
amplitude and the TFF, we choose twisted basis operators which correspond to the physical-basis electromagnetic currents and pion operators. Specifically, we choose the pion annihilation operators to be
\begin{equation}
  \begin{aligned}
    P^0(x) &= i\,\psibar(x) \gamma_5 \tau^3 \psi(x)
    = -\chibar(x) \chi(x), \\
    P^{\pm}(x) &= i\, \psibar(x) \gamma_5 \tau^{\pm} \psi(x)
    = i \, \chibar(x) \gamma_5 \tau^\pm \chi(x),
  \end{aligned}
  \label{eq:ps-operators}
\end{equation}
where $\tau^{\pm}$ are the raising/lowering operators of isospin.
We also choose local electromagnetic currents, given in the physical basis in the light-quark sector by
\begin{equation}
  j^{\,(l)}_\mu(x) = \psibar(x) \,\gamma_\mu \, Q^{(l)} \, \psi(x),
\end{equation}
where $Q^{(l)} = \mathrm{diag}(2/3, \, -1/3)$ is the light-quark charge matrix.
This current can be decomposed into terms of total isospin $I=1$ and $I=0$ (always with fixed $z$-component $I_z = 0$), denoted respectively by $j^{1,0}$ and $j^{0,0}$. The result is then easily written in the twisted basis, giving
\begin{equation}
\begin{aligned}
  j^{\,(l)}_\mu(x) &= \frac{1}{6} \, j^{\,0,0}_\mu(x) + \frac{1}{2} \, j^{1,0}_\mu(x) \\
  &= \frac{1}{6}\, \chibar(x) \gamma_\mu \chi(x) + \frac{1}{2} \, \chibar(x) \gamma_\mu \tau^3 \chi(x).
\end{aligned}
\end{equation}
It will also be useful for the following section to define the currents with isospin $I=1$ and $I_z = \pm1$,
\begin{equation}
  j^{1,\pm}_\mu(x) = \psibar(x) \gamma_\mu \tau^{\pm} \psi(x)
  = \mp i \, \chibar(x) \gamma_\mu \gamma_5 \tau^{\pm} \chi(x).
\end{equation}

For the strange and charm quark we use a mixed-action approach: 
In the sea-quark action the strange and charm quark
form a  twisted-mass quark doublet, including tuned twisted-mass and quark-mass splitting terms.
Details about tuning of the twisted-mass parameters for the sea-quark
action, in particular to maximal twist, are given in Ref.~\cite{Alexandrou:2018egz}.
In the valence action we
use two Osterwalder-Seiler heavy-quark doublets \cite{Frezzotti:2004wz} by introducing
two doublets $\chi^{(h)} = (\chi^{(h)}_{+}, \, \chi^{(h)}_{-})$, for $h = s,\,c$, which
avoids twisted-mass-related mixing of strange and charm flavor.
Both doublets work analogously to the up- and down-quark doublet,
 they have the same critical
hopping parameter, but use individual tuning of the twisted quark-mass
parameters. The strange-quark mass 
is tuned such that the $\Omega$
baryon takes its physical value, while the charm-quark mass is tuned to reproduce a near-physical $\Lambda_c$ mass.

The strange- and charm-flavor electromagnetic currents then use the local operators 
\begin{align}
  j^{\,(s)} &= -\frac{1}{3}\,  \psibar^{(s)}(x) \, \gamma_\mu \, \psi^{(s)}(x) 
    = -\frac{1}{3} \, \chibar^{(s)}(x) \, \gamma_\mu \, \chi^{(s)}(x)\,,
    \nonumber \\
    j^{\,(c)} &= \phantom{-}\frac{2}{3} \, \psibar^{(c)}(x) \, \gamma_\mu  \, \psi^{(c)}(x)
    = \phantom{-} \frac{2}{3} \, \chibar^{(c)}(x) \, \gamma_\mu \, \chi^{(c)}(x)\,.
    \label{eq:sc-em-current}
\end{align}

The iso-scalar strange- and charm-flavor currents enter correlation functions in \Fig~\ref{fig:wick_contractions} only
as vector-current disconnected diagrams ``V-disconnected I'' and ``V-disconnected II'' (details on the Wick contractions are provided in the following section). To maintain automatic 
$\mathcal{O}(a)$ improvement we average the vector-current disconnected contributions from strange and charm propagators
generated from both $\chi^h_{+}$ and $\chi^{h}_{-}$ with $h = s,\,c$.

We use gauge ensembles produced by the Extended Twisted Mass Collaboration (ETMC) from
simulations of isospin-symmetric QCD (isoQCD) with $N_f = 2+1+1$ flavors of Wilson Clover twisted mass quarks.
The properties of the used ensembles most relevant for this work 
can be found in \Tab~\ref{table:ensemble_params};
for more  details see Refs.~\cite{Alexandrou:2018egz,Bergner:2020vQ,ExtendedTwistedMass:2021qui,Finkenrath:2022Q3,ExtendedTwistedMass:2022jpw}.
\setlength\tabcolsep{10pt} 
\begin{table*}[!th]
     \centering
          \begin{tabular}{|| c c c c c c || c || c c ||}
          \hline
          Ensemble & $V/a^4$ & $N_{\mathrm{conf}}$ & $\beta$ & $a$\,[fm] & $a \mu_l$ & $m_\pi$\,[MeV] & $L$\,[fm] & $m_\pi L$\\ \hline
          cB211.072.64 & $64^3 \cdot 128$ & 748 & 1.778 & 0.07961(13) & 0.00072 & 140.2(2) & 5.09 & 3.62 \\
          cC211.060.80 & $80^3 \cdot 160$ & 397 & 1.836 & 0.06821(12) & 0.00060 & 136.7(2) & 5.46 & 3.78 \\
          cD211.054.96 & $96^3 \cdot 192$ & 495 & 1.900 & 0.05692(10) & 0.00054 & 140.8(2) & 5.46 & 3.90 \\
          \hline
     \end{tabular}
     	\caption{Parameters of the ETMC ensembles for the analysis presented in this work, adapted from \cite{ExtendedTwistedMass:2022jpw}. Further parameters can be found in \cite{Alexandrou:2018egz,Bergner:2020vQ,ExtendedTwistedMass:2021qui,Finkenrath:2022Q3,ExtendedTwistedMass:2022jpw}. 
     	\label{table:ensemble_params}}
\end{table*}

The local vector currents detailed above renormalize multiplicatively.
Because of the twisted-mass formulation, iso-scalar and neutral iso-vector vector currents are renormalized with renormalization factor $Z_V$,
while the charged iso-vector vector currents renormalize with $Z_A$:
\begin{align}
j_{\mu, R}^{\,0,0} &= Z_V j_{\mu}^{\,0,0}\,,\quad 
j_{\mu, R}^{1,0} = Z_V j_{\mu}^{1,0}\,, \\
j_{\mu, R}^{1,\pm} &= Z_A j_{\mu}^{1,\pm}\,, \\
j_{\mu, R}^{\,(s)} &= Z_V j_{\mu}^{\,(s)}\,, \quad
j_{\mu, R}^{\,(c)} = Z_V j_{\mu}^{\,(c)} \,.
\end{align}
These renormalization constants (RCs) were calculated to high precision for the ensembles used in this work in Ref.~\cite{ExtendedTwistedMass:2022jpw}, as shown in \Tab~\ref{table:renorm_consts}.
The uncertainties on the RCs are negligible compared to the errors on the bare TFFs, such that we only use the central values of the RCs throughout this work. 

	\begin{table}[!h]
	\centering
	\begin{tabular}{||c | c c||} 
		\hline
		Ensemble & $Z_V$ & $Z_A$ \\ 
		\hline
		cB211.072.64 & 0.706379(24) & 0.74294(24) \\ 
		\hline
		cC211.060.80 & 0.725404(19) & 0.75830(16) \\ 
		\hline
		cD211.054.96 & 0.744108(12) & 0.77395(12) \\ 
		\hline
	\end{tabular}
	\caption{Values of $Z_V$ and $Z_A$ for the ETMC ensembles which were used in this work, as determined in Ref.~\cite{ExtendedTwistedMass:2022jpw} using a hadronic method.
    \label{table:renorm_consts}}
\end{table}
\setlength\tabcolsep{5pt} 

Two remarks are in order due to the fact that we are using 
nonconserved local vector currents for the construction of the three-point
correlation function in \Eq~\eqref{eq:amplitude}. First, one can show that possible short-distance singularities
are absent, i.e., the local currents in \Eq~\eqref{eq:amplitude} admit
a well defined continuum limit. The arguments are based on the
operator product expansion and are given in Appendix D of
Ref.~\cite{Gerardin:2016cqj} for Wilson fermions. Due to universality,
they also apply to the Wilson twisted-mass lattice QCD
formulation. Second, we note that the nonconserved local currents do
not spoil the automatic $O(a)$ improvement. This can be seen from the
fact that all involved lattice quantities are constructed such that
they have the correct symmetry property under the combined twisted-mass
transformation ${\cal P} \times {\cal D}_d \times (\mu \to -\mu)$
where ${\cal P}$ is the standard parity symmetry transformation, ${\cal D}_d$
the operator-dimensionality transformation, and $\mu$ the
twisted-quark mass parameter. This symmetry forbids the appearance of
$O(a)$ terms in physical matrix elements for twisted-mass lattice QCD at maximal
twist  \cite{Frezzotti:2003ni, Frezzotti:2004wz}, and hence guarantees automatic $O(a)$-improvement of the three-point function in \Eq~\eqref{eq:amplitude}.

\subsection{Wick contractions}

A naive evaluation of the three-point amplitude
\Eq~\eqref{eq:amplitude} contains disconnected contributions in which
the quarks within the neutral pion contract with themselves. These
include a singly-disconnected contribution which correlates the
neutral-pion loop with the connected vector-vector two-point function,
as well as a doubly-disconnected contribution with three
loops. In the limit of exact isospin
symmetry, these contributions vanish
identically. In the twisted-mass formulation at finite lattice
spacing, however, isospin symmetry is broken by $\mathcal{O}(a^2)$
lattice artifacts~\cite{Frezzotti:2000nk} (independently of the twist
angle) and the disconnected contributions do not cancel at finite lattice spacing.

\begin{figure}[t!]
    \centering
    \newcommand{\tikzTickSize}[0]{2pt}
\newcommand{\tikzJmuX}[0]{1.0}
\newcommand{\tikzJmuY}[0]{1.0}
\newcommand{\tikzJnuX}[0]{2.0}
\newcommand{\tikzAxisY}[0]{-0.5}
\newcommand{\tikzAxisStart}[0]{-0.5}
\newcommand{\tikzAxisEnd}[0]{2.5}
\newcommand{\tikzPX}[0]{0.0}
\newcommand{\tikzPloopX}[0]{0.5}
\newcommand{\tikzTitleY}[0]{1.6}
\newcommand{\tikzFermBend}[0]{25}
\newcommand{\tikzTitleSize}[0]{\footnotesize}
\newcommand{\tikzAxLabelSize}[0]{\scriptsize}
\newcommand{\tikzLabelSize}[0]{\scriptsize}

\tikzfeynmanset{
  cross/.style={/tikz/path picture={ 
    \draw[black]
    (path picture bounding box.south east) -- (path picture bounding box.north west) (path picture bounding box.south west) -- (path picture bounding box.north east);
  }},
  current/.style={
    /tikz/circle,
    /tikz/fill=white,
    /tikz/draw=black,
    /tikz/inner sep=0.7mm,
    /tikzfeynman/cross
  },
  ps/.style={
    /tikz/circle,
    /tikz/draw=black,
    /tikz/fill,
    /tikz/inner sep=0.5mm
  },
  tiny/.style={
    /tikzfeynman/arrow size=0.8pt
  }
}

\newcommand{\wickDiagram}[2]{
\begin{tikzpicture}[scale=1.3]

#2

\begin{feynman}[tiny]

\vertex [ps, label={[inner sep=0]south west:\tikzLabelSize \( P^{\pm \dag} \)}] (p) at (\tikzPX,0) {};
\vertex [current, label={[inner sep=0]north west:\tikzLabelSize \( j_\nu \)}] (mu) at (\tikzJmuX,\tikzJmuY) {};
\vertex [current, label={[inner sep=0]south east:\tikzLabelSize \( j_\mu \)}] (nu) at (\tikzJnuX,0) {};
\vertex (ploop) at (\tikzPloopX,0);

\diagram* {
#1
};

\end{feynman}

\draw[-latex'] (\tikzAxisStart,\tikzAxisY) -- (\tikzAxisEnd,\tikzAxisY);
\draw ($(\tikzPX,\tikzAxisY) + (0,-\tikzTickSize)$) -- ($(\tikzPX,\tikzAxisY) + (0,\tikzTickSize)$) node [below = 1mm] {\tikzAxLabelSize $-t_{\pi}$};
\draw ($(\tikzJmuX,\tikzAxisY) + (0,-\tikzTickSize)$) -- ($(\tikzJmuX,\tikzAxisY) + (0,\tikzTickSize)$) node [below = 1mm] {\tikzAxLabelSize $0$};
\draw ($(\tikzJnuX,\tikzAxisY) + (0,-\tikzTickSize)$) -- ($(\tikzJnuX,\tikzAxisY) + (0,\tikzTickSize)$) node [below = 1mm] {\tikzAxLabelSize $\tau$};

\end{tikzpicture}
}

\newcommand{\drawPloop}[0]{
\draw [/tikzfeynman/tiny,postaction={/tikzfeynman/with arrow=0.4}] (p) arc (180:540:1.8mm);
}
\newcommand{\drawJmuloop}[0]{
\draw [/tikzfeynman/tiny,postaction={/tikzfeynman/with arrow=0.4}] (mu) arc (90:450:1.8mm);
}
\newcommand{\drawJnuloop}[0]{
\draw [/tikzfeynman/tiny,postaction={/tikzfeynman/with arrow=0.4}] (nu) arc (0:360:1.8mm);
}

\wickDiagram{(p) -- [fermion] (nu) -- [fermion] (mu) -- [fermion] (p)}{
\node at (\tikzJmuX,\tikzTitleY) {\tikzTitleSize Connected I};
}
\wickDiagram{(p) -- [fermion] (mu) -- [fermion] (nu) -- [fermion] (p)}{
\node at (\tikzJmuX,\tikzTitleY) {\tikzTitleSize Connected II};
} \\
\wickDiagram{(p) -- [fermion, bend right=\tikzFermBend] (mu) -- [fermion, bend right=\tikzFermBend] (p)}{
\drawJnuloop
\node at (\tikzJmuX,\tikzTitleY) {\tikzTitleSize V-disconnected I};
}
\wickDiagram{(p) -- [fermion, bend right=\tikzFermBend] (nu) -- [fermion, bend right=\tikzFermBend] (p)}{
\drawJmuloop
\node at (\tikzJmuX,\tikzTitleY) {\tikzTitleSize V-disconnected II};
}
    \caption{Wick contractions contributing to $C_{\mu\nu}^\pm(\tau,t_P)$. 
    These are the connected (top row), and vector-current disconnected (``V-disconnected'', bottom row) diagrams. Note that the connected diagrams only involve light quarks, while the disconnected vector-current loop in the V-disconnected diagrams involves light, strange, and charm quarks for the $2+1+1$ ensembles used in this work.
    }
    \label{fig:wick_contractions}
\end{figure}
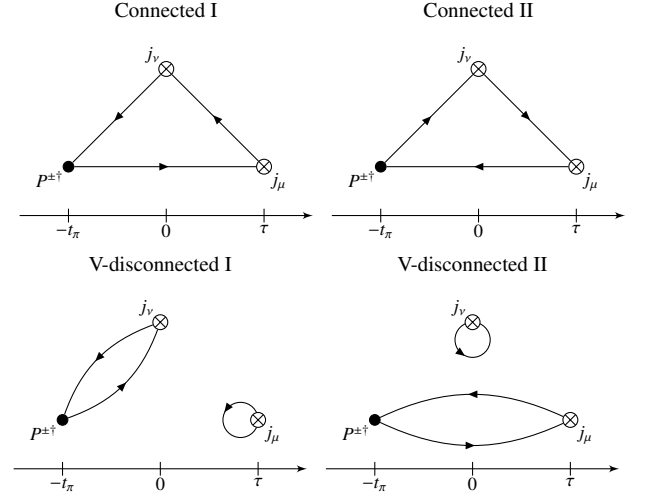

Since typically disconnected diagrams are costly to evaluate, we minimize the number of such diagrams by instead considering the isospin-rotated amplitude
\begin{equation}
\begin{aligned}
\label{eq:isospin_preserving_amplitude_pm}
C_{\mu \nu}' &= -\frac{1}{6} \left\langle j_\mu^{\,0,0} j_\nu^{1,-} P^+  \right\rangle + \frac{1}{6} \left\langle j_\mu^{\,0,0} j_\nu^{1,+} P^- \right\rangle \\
&\qquad + \frac{1}{3} \left< j_\mu^{\,(s)} j_\nu^{1,-} P^+ \right> - \frac{1}{3} \left< j_\mu^{\,(s)} j_\nu^{1,+} P^- \right> \\
&\qquad - \frac{2}{3} \left< j_\mu^{\,(c)} j_\nu^{1,-} P^+ \right> + \frac{2}{3} \left< j_\mu^{\,(c)} j_\nu^{1,+} P^- \right>
\end{aligned}
\end{equation}
where spacetime coordinates have been suppressed for clarity.
The amplitude in \Eq~\eqref{eq:isospin_preserving_amplitude_pm} does not contain pion-loop disconnected contributions, 
leaving only the connected and vector-current disconnected
(``V-disconnected'') contributions depicted in
\Fig~\ref{fig:wick_contractions}.
This rotation is valid in the continuum due to the iso-symmetric continuum limit of twisted-mass lattice QCD.
Consequently, the rotation only modifies the lattice artifacts at $O(a^2)$ in the twisted-mass formulation, which is taken into
account in the continuum extrapolation.

We evaluate the connected diagrams using inversions based on point sources for the current $j_\nu(\vec{0},0)$ with sequential inversion through the pseudoscalar operator $P^{\pm \dag}$. This means that each choice of momentum $\vec{p}$ for the pseudoscalar requires a new inversion, while momentum $\vec{q}_1$ inserted at the sink current $j_\mu(\vec{x},\tau)$ can be easily varied. 
For the evaluation of the connected contribution to the three-point amplitude, 16 point sources per configuration are used for cB211.072.64 
and 8 point sources per configuration on the other two ensembles cC211.06.80 and cD211.054.96 in \Tab~\ref{table:ensemble_params}. 

For the  vector-current disconnected diagrams ``V-disconnected I/II'' we use quark loops generated with hierarchical probing 
\cite{Stathopoulos:2013aci} in general, and low-mode deflation \cite{Gambhir:2016uwp} in particular on ensemble cB211.072.64.
Details on the dilution scheme for the Hadamard vectors and the number of low modes are given in Ref.~\cite{Alexandrou:2020sml}
and in Table 2 of Ref.~\cite{Alexandrou:2022osy}. The associated two-point correlators built from the charged pion
and iso-vector vector current (cf.\ \Fig~\ref{fig:wick_contractions}) are evaluated with stochastic timeslice propagators
based on $\mathbb{Z}_2 \times i\, \mathbb{Z}_2$ noise sources. The number of stochastic samples is chosen equal to the time extent in lattice units, $T/a$.

\subsection{Kinematics}

\begin{figure*}[t!]
	\centering
	\includegraphics{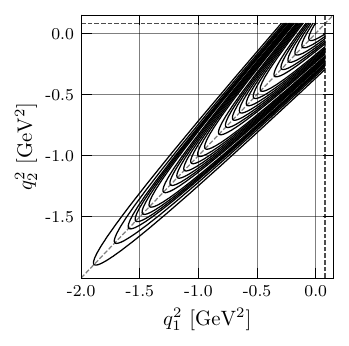}
	\includegraphics{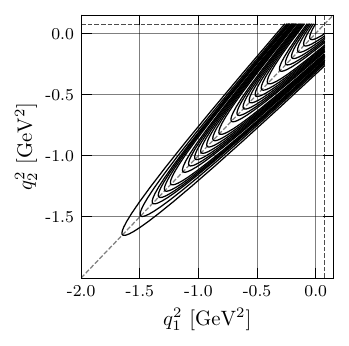}
	\includegraphics{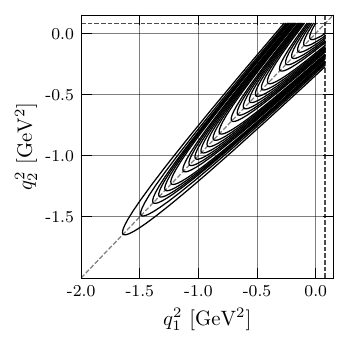}
	\caption{Range of photon virtualities spanned in our calculation on the ensembles cB211.072.64 (left), cC211.060.80 (middle), and cD211.054.96 (right). Dotted lines at $q_{1,2}^2 = 4 m_\pi^2$ indicate the nonanalytic threshold beyond which the rotation to Euclidean spacetime is no longer valid. 
    \label{fig:yield}}
\end{figure*}
The pion momentum $\vec p$ is set through the pseudoscalar
interpolating operator while the momentum $\vec{q}_1$ is set through
the current $j_\mu$, as shown in \Eq~\eqref{eq:amplitude}, with both
momenta selected from the available finite-volume momenta
$\vec{q}_1,\vec{p} = 2 \pi/L \cdot \vec n$, where $\vec{n} \in \mathbb{Z}^3$.  Momentum conservation sets the momentum $\vec{q}_2 = \vec{p} - \vec{q}_1$. The energy $E_\pi$ of the pion is then imposed by the on-shell condition, while 
the temporal component $\omega_1$ of $q_1$ can be continuously varied as indicated in \Eq~\eqref{eq:matelem_eucl}, with the temporal component $\omega_2$ of $q_2$ fixed by energy conservation to $\omega_2 = E_\pi - \omega_1$.
The kinematical range accessible on the lattice can be parameterized by the accessible virtualities of the electromagnetic currents, 
\begin{align}
\label{eq:virtualities}
q_1^2 &= \omega_1^2 - \vec q_1^{\,2}, \nonumber \\ 
q_2^2 &= (E_\pi - \omega_1)^2 - (\vec p - \vec q_1)^2.
\end{align}
For the Wick rotation to the Euclidean metric to be valid, the virtualities must fall below the two-pion threshold, i.e., $q_i^2 < 4 m_\pi^2$.

For a pion at rest, $\vec p = 0$, it holds that $E_\pi = m_\pi$ and $\vec q_2 = -\vec q_1$. In this case, the photon virtualities simplify to 
\begin{align}
\label{eq:virtualities_p0}
q_1^2 &= \omega_1^2 - \vec q_1^{\,2}, \nonumber \\ 
q_2^2 &= (m_\pi - \omega_1)^2 - \vec q_1^2.
\end{align}
The analyticity constraint becomes a constraint on $\omega_1$,
\begin{equation}
\sqrt{4 m_\pi^2+\vec q_1^2} + m_\pi < \omega_1 < \sqrt{4 m_\pi^2+\vec q_1^2}.
\end{equation}
\Fig~\ref{fig:yield} depicts the kinematical range for 26 different choices of $|\vec{q}_1|^2$ satisfying $(2\pi/L)^2 \leq |\vec{q}_1|^2 \leq 32 (2\pi/L)^2$ for each of the finite volumes accessible on the three distinct ensembles given in \Tab~\ref{table:ensemble_params}.
For all three ensembles, the maximum momentum used in the evaluation of the three-point correlation function gives access to virtualities up to $|q_{1,2}^2| \approx 1.7\,\text{GeV}^2$. 

Using
\Eqs~\eqref{eq:minkowskiFF},~\eqref{eq:matelem_eucl}~and~\eqref{eq:virtualities_p0}
it is straightforward to show that in the rest frame of the pion $\tilde{A}_{\mu \nu}(\tau)$ vanishes when one or both Lorentz indices are temporal, while the spatial components can be written as
\begin{equation}
\label{eq:A_tilde_def}
\tilde{A}_{ij}(\tau) = -i m_\pi \epsilon_{ijk} q_1^k \tilde{A}(\tau).
\end{equation}
Inverting the relation gives
\begin{equation}
\label{eq:A_tilde_scalar}
\tilde{A}(\tau) = i \epsilon_{ijk} \frac{q_1^k}{m_\pi \vec q_1^{\,2}} \tilde{A}_{ij}(\tau),
\end{equation}
where $\tilde A (\tau)$ is a scalar under the spatial rotation group.
Combining \Eqs~\eqref{eq:minkowskiFF}, \eqref{eq:matelem_eucl} and \eqref{eq:A_tilde_scalar} shows that the form factor can be extracted from a Laplace transform of this scalar quantity as
\begin{equation} \label{eq:TFF_from_A}
    \Fpigg(q_1^2, q_2^2) = \int_{-\infty}^\infty d\tau \, \tilde A (\tau) e^{\omega_1 \tau}.
\end{equation}
Since the full TFF can be extracted from the scalar $\tilde A(\tau)$
in the rest frame of the pion, we focus on the evaluation of this scalar function for the remainder of this work.

The scalar amplitude $\tilde{A}(\tau)$ is implicitly parameterized by the momentum $\vec{q}_1$. All choices of momenta with a fixed value of $|\vec{q}_1|^2$ can be used to evaluate the form factor as in \Eq~\eqref{eq:TFF_from_A} at identical kinematics. We therefore choose to average the scalar amplitude over such equivalent momenta to increase statistics.

\subsection{Finite-time extent corrections}
All ensembles in this work use periodic (anti-periodic) temporal boundaries for the fundamental bosons (fermions). As a result, the pion experiences a periodic boundary condition in time, and the three-point function $C_{\mu\nu}$ receives corrections from states propagating backwards around the finite-time extent of our lattices.
Denoting the lattice time extent $T$, including the leading correction gives
\begin{equation}
  \begin{aligned}
    C_{\mu\nu}(\tau, t_\pi) &\approx e^{-E_\pi t_\pi} \frac{Z_\pi}{2 E_\pi} \left< 0 \left| T \{ j_\mu(\tau) j_\nu(0) \} \right| \pi^0 (\vec{p}) \right> \\
    &+ e^{-E_\pi (T - t_\pi)} \frac{Z_\pi}{2 E_\pi} \left< \pi^0(\vec{p}) \left| T \{ j_\mu(\tau) j_\nu(0) \} \right| 0 \right> \\
    &+ O(e^{-\Delta E \tsep} + e^{-\Delta E (T - |\tau| - \tsep)}). \\
  \end{aligned}
\end{equation}

In the rest frame of the pion, we can apply Bose symmetry and PT symmetry to define a corrected scalar amplitude $\tilde{A}(\tau)$ (cf.~\Eq~\eqref{eq:A_tilde_scalar}) to account for this leading-order effect,
\begin{equation} \label{eq:A_tilde_ft}
  \begin{aligned}
    \tilde{A}(\tau; t_\pi) &\equiv i \epsilon_{ijk} \frac{q_1^k}{m_\pi \vec{q}_1^2} \left[1 - e^{E_{\pi} (T - |\tau| - t_\pi)} \right]^{-1} \tilde{A}_{ij}(\tau; t_\pi) \\
    &= \tilde{A}(\tau) + O(e^{-\Delta E \tsep} + e^{-\Delta E (T - |\tau| - \tsep)}).
  \end{aligned}
\end{equation}
In all further analysis, this corrected form of the lattice amplitude is used as input.

\begin{figure*}[t!]
	\centering
	\includegraphics{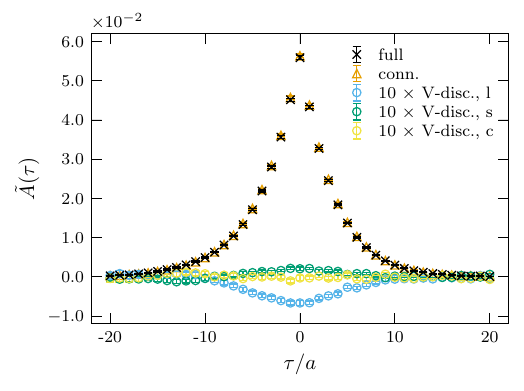}
	\includegraphics{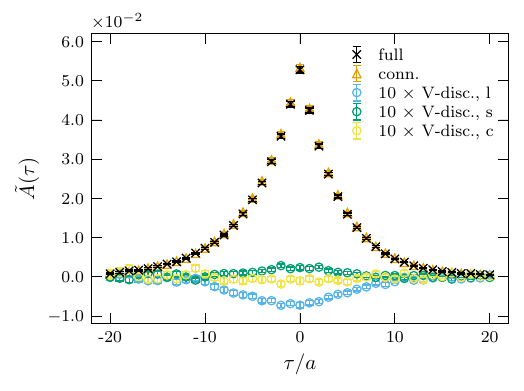}
	\caption{$\tilde A (\tau)$ 
 with $|\vec q_1|^2 = 10 (2 \pi/L)^2$ including finite-time extent
 corrections on the ensembles cC211.060.80 (left) and cD211.054.96
 (right). We show the full $\tilde A (\tau)$ as black crosses, the
 connected contribution as orange triangles, and the V-disconnected
 contributions (multiplied by 10 for easier comparison) with a light,
 strange and charm vector-current loop as blue, green and yellow circles, respectively. \label{fig:A_tilde}
      }
\end{figure*}

\subsection{Dependence on \texorpdfstring{$t_\pi$}{t\_pi}}
\begin{figure*}[t!]
	\centering
	\includegraphics{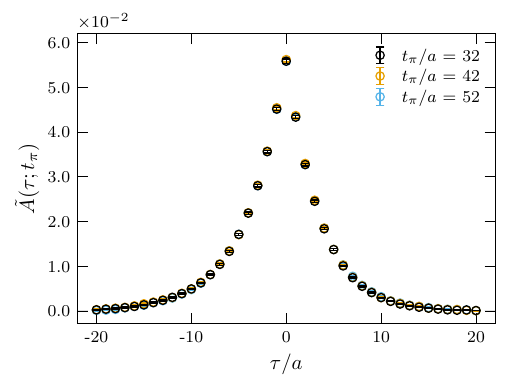}
	\includegraphics{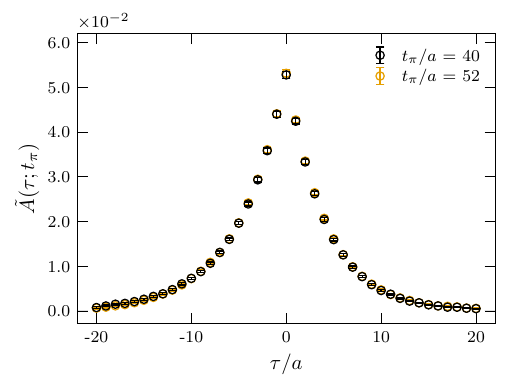}
	\caption{$\tilde A (\tau; t_\pi)$ 
 with $|\vec q_1|^2 = 10 (2 \pi/L)^2$  including finite-time extent corrections on the ensembles cC211.060.80 (left) and cD211.054.96 (right) for the choices $t_\pi \in \{2.18, 2.86, 3.55\}\,$fm (left) and $t_\pi \in \{2.28, 3.02\}\,$fm (right). \label{fig:A_tilde_tpi}
      }
\end{figure*}

An example of the scalar amplitude $\tilde{A}(\tau)$ with the decomposition into contributions from individual Wick contractions is given in \Fig~\ref{fig:A_tilde}. These results, along with those used in the following analysis are based on the choice of interpolating operator time $t_\pi = 2.23\,$fm for ensemble cB211.072.64, $t_\pi = 2.18$\,fm for ensemble cC211.060.80, and $t_\pi = 2.28$\,fm for ensemble cD211.054.96, with the choices made such that $t_\pi$ in physical units is similar on all three ensembles. Note that smaller values of $t_\pi$ lead to a reduction of the statistical errors, however, this requires a careful analysis of excited state effects. 
To confirm that our choices do not suffer from significant excited state effects, two larger choices of $t_\pi$ were measured on cB211.072.64 and cC211.060.80 and one larger choice on cD211.054.96. We depict an example of the comparison of $\tilde{A}(\tau;t_\pi)$ across different choices of $t_\pi$ in \Fig~\ref{fig:A_tilde_tpi}. In all cases, $\tilde{A}(\tau;t_\pi)$ does not show any systematic dependence on $t_\pi$ across the choices.

\subsection{Tail Fits}
The extraction of the transition form factor using \Eq~(\ref{eq:matelem_eucl}) in principle requires access to $\tilde{A}(\tau)$ for arbitrary $\tau \in (-\infty, \infty)$.
To estimate the amplitude over the whole temporal range from our lattice data with finite temporal extent, we use two phenomenological models, the Vector Meson Dominance model (VMD) and Lowest Meson Dominance model (LMD)~\cite{PhysRevD.94.053006,PhysRevD.51.4939,PhysRevLett.83.5230}.
The VMD and LMD form factors are given, respectively, by
\begin{equation}
\Fpigg^{\mathrm{VMD}}(q_1^2, q_2^2) = \frac{\alpha M_V^4}{(M_V^2 - q_1^2) (M_V^2 - q_2^2)}
\end{equation}
and
\begin{equation}
\label{eq:FF_LMD}
\Fpigg^{\mathrm{LMD}}(q_1^2, q_2^2) = \frac{\alpha M_V^4 + \beta (q_1^2 + q_2^2)}{(M_V^2 - q_1^2) (M_V^2 - q_2^2)}.
\end{equation}
The phenomenological origin of the models suggests particular parameter choices. In the chiral limit and at low energy, the fermionic triangle diagram gives rise to the Adler-Bell-Jackiw (ABJ) anomaly \cite{PhysRev.177.2426,BellJackiw},
which constrains the form factor in these limits to 
\begin{equation}
\label{eq:ABJ}
\Fpigg (0,0) = \frac{1}{4 \pi^2 F_\pi},
\end{equation}
with $F_\pi = 92.3(1)$\,MeV the pion decay constant \cite{Workman:2022ynf}, suggesting $\alpha = 1/(4 \pi^2 F_\pi) = 0.274\,\mathrm{GeV}^{-1}$. Meanwhile, the mass $M_V = 775\,\mathrm{MeV}$ is the $\rho$-meson mass in the VMD and LMD models, and 
$\beta = - F_\pi/3 = -0.0308\,\mathrm{GeV}$
reproduces the leading OPE prediction \cite{PhysRevD.51.4939,PhysRevLett.83.5230,Gerardin:2016cqj}. However, we treat $\alpha, M_V$ and $\beta$ as free model parameters when fitting the models to our data, using the models as inspiration for the fit form rather than a prediction of the precise asymptotic behavior.

Using the relation between the TFF and amplitude in \Eq~\eqref{eq:minkowskiFF} and inverting the relation in \Eq~\eqref{eq:matelem_eucl}, the VMD and LMD form factors can be converted into fit functions for the Euclidean-time amplitude $\tilde{A}$ as follows.
For the LMD fit function one finds 
\begin{equation}
\begin{aligned}
\label{eq:A_tilde_fitfunc}
    \tilde{A}^{\mathrm{LMD}}(\tau) &= 
    e^{m_\pi |\tau| \Theta(-\tau)} [ C_+ e^{-E_V |\tau|}
    - C_- e^{-(m_\pi + E_V)|\tau|}],
\end{aligned}
\end{equation}
in terms of
\begin{equation}
  \begin{aligned}
    C_{\pm} &\equiv \frac{\alpha M_V^4 + \beta (2 M_V^2 + m_\pi^2 \mp 2 m_\pi E_V)}{m_\pi E_V(2E_V \mp m_\pi)}, \\
    E_V &\equiv \sqrt{M_V^2+|\vec{q}_1|^2}.
  \end{aligned}
\end{equation}
The VMD fit function is obtained by setting $\beta = 0$.

The VMD and LMD models should not be expected to fully capture the large-$Q^2$ behavior of the form factor, or conversely the small-$|\tau|$ behavior of $\tilde{A}(\tau)$. In particular,
the Brodsky-Lepage behavior constrains the single-virtual form factor at large Euclidean (space-like) momentum \cite{Lepage:1979zb,Lepage:1980fj,Brodsky:1981rp}, giving 
\begin{equation}
\label{eq:Brodsky-Lepage}
\Fpigg (-Q^2,0) \overset{Q^2 \rightarrow \infty}{\longrightarrow} \frac{2 F_\pi}{Q^2}
\end{equation}
at leading order in $\alpha_s$.
This behavior is reproduced by the VMD model, but the LMD model tends to a constant at large Euclidean momenta in the single-virtual form factor. 
Meanwhile, the operator product expansion (OPE) at short distances \cite{Nesterenko:1982dn,Novikov:1983jt} restricts the double-virtual form factor where both momenta become large at the same time, giving in the chiral limit
\begin{equation}
\label{eq:OPE}
\Fpigg (-Q^2,-Q^2) \overset{Q^2 \rightarrow \infty}{\longrightarrow} \frac{2 F_\pi}{3}\left[\frac{1}{Q^2} + \mathcal{O}\left(\frac{1}{Q^4}\right) \right].
\end{equation}
This behavior is reproduced by the LMD model (with the choice of $\beta$ given above), but the VMD model in this case falls off as $1/Q^4$ \cite{Gerardin:2016cqj}.
Despite these shortcomings at large $Q^2$, both models accurately capture the long-distance behavior. As such, we use the lattice data directly for the short-distance region while replacing only the tails with these fits. Details of this procedure are discussed in the following section.

\section{Transition Form Factor}
\label{sec:TFF}

To obtain the transition form factors, 
we need to evaluate the Laplace transform in \Eq~\eqref{eq:TFF_from_A}, reproduced here for convenience:
\begin{equation}
\Fpigg(q_1^2, q_2^2) = \int_{-\infty}^\infty d\tau \, \tilde A (\tau)
e^{\omega_1 \tau},
\label{eq:TFF_from_A_2}
\end{equation}
where for a given momentum orbit $|\vec q_1|^2$ the squared
four-momenta $q_1^2$ and $q_2^2$ are given by
\Eq~\eqref{eq:virtualities_p0}. The TFF is only directly available on
specific sets of orbits $q_{1,2}^2$ determined by the kinematics, and
it must therefore be interpolated/extrapolated to other points in the
$(q_1^2, q_2^2)$ plane. In the following we describe our strategy for
the numerical integration in \Eq~\eqref{eq:TFF_from_A_2} and for the
extension of the TFF lattice data to the whole momentum plane using
the modified $z$-expansion.

\subsection{Numerical integration}
\label{ssec:NumericalIntegration}

\begin{figure*}[th!]
	\centering
	\includegraphics[page=6]{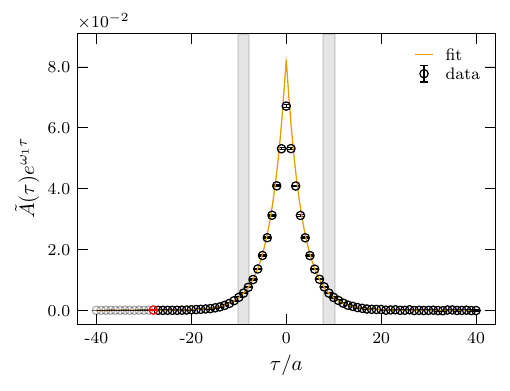}
	\includegraphics[page=6]{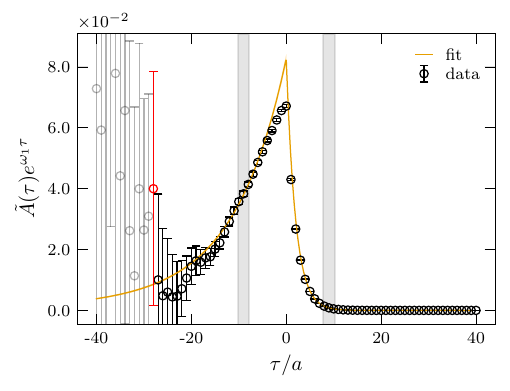}
	\caption{Integrands $\tilde A (\tau) e^{\omega_1 \tau}$ 
 with $|\vec q_1|^2 = 6 (2 \pi/L)^2$ on the ensemble cB211.072.64, comparing diagonal kinematics $q_1^2 = q_2^2$ (left) and single-virtual kinematics $q_2^2 = 0$ (right). 
	We show the integrand from lattice data as black circles and
        from a correlated LMD fit to $\tilde{A}(\tau)$ according to
        \Eq~\eqref{eq:A_tilde_fitfunc} as an orange line. The fit was
        done globally including all momentum orbits $(2\pi/L)^2 \leq
        |\vec{q}_1|^2 \leq 32 (2\pi/L)^2$ with fit range $[-10, -8]
        \cup [8, 10]$, indicated by the grey bands, yielding
        $\chi^2/\mathrm{d.o.f.} = 1.01$ with correlations taken into
        account across all orbits and fit range. The red data point
        indicates the timeslice where the pseudoscalar operator is
        inserted, while the greyed-out points are those where the time
        ordering does not satisfy the one from \Eq~\eqref{eq:tilde_A_timeordering}. 
	\label{fig:integrand}}
    \end{figure*}

The tails of $\tilde A (\tau)$ associated with large $|\tau|$ are expected to decrease exponentially quickly, with a mass scale predicted by the VMD/LMD models to be $E_V = \sqrt{M_V^2 + |\vec{q}_1|^2}$ in one tail and $E_V + m_\pi$ in the other (see \Eq~\eqref{eq:A_tilde_fitfunc}).
However, the exponential factor $e^{\omega_1 \tau}$ enhances one tail of $\tilde A (\tau)$ significantly whenever $|\omega_1|$ is sufficiently large compared to this energy scale. This makes the numerical integration difficult: Since the signal-to-noise ratio decreases when going to larger $|\tau|$, fluctuations in one of the tails due to the small signal-to-noise ratio are exponentially enhanced. We observe that this effect is small for virtualities near the diagonal $q_1^2 = q_2^2$, for which $\omega_1 = m_{\pi}/2 \ll E_V$,
but can be severe in the single-virtual cases corresponding to either $q_1^2 = 0$ or $q_2^2 = 0$,  for which $\omega_1 = |\vec{q}_1| \approx E_V$,
especially for the larger momentum orbits. \Fig~\ref{fig:integrand}
    shows a representative comparison of the integrand of the Laplace transform for both the diagonal and single-virtual cases.
For choices of virtualities between these two cases, the exponential enhancement gets more and more pronounced when going from the diagonal to the single-virtual case.

These issues are addressed by the introduction of cutoff times $\tau^{R}_{\mathrm{cut}}$ and $\tau^{L}_{\mathrm{cut}}$, with lattice data replaced by a global fit to $\tilde A (\tau)$ for $\tau \geq \tau^{R}_{\mathrm{cut}}$ and $\tau \leq \tau^{L}_{\mathrm{cut}}$.
In our analysis, we perform a fit to \Eq~\eqref{eq:A_tilde_fitfunc}
across data at all available $\vec{q}_1$ and for values of $\tau$
selected from symmetrical fit ranges on both sides of the peak, i.e.,
$\tau \in [-\tau_{\mathrm{max}}, -\tau_{\mathrm{min}}] \cup
[\tau_{\mathrm{min}}, \tau_{\mathrm{max}}]$. Variation of the fit
ranges allows us to determine a corresponding systematic error.
The left and right cutoff times are selected depending on the sign of $\omega_1$ to include as much data as possible on the exponentially suppressed tail of the integrand, in particular by fixing
$\tau^{L}_{\mathrm{cut}} = -T/2$ for $\omega_1 \geq 0$ or $\tau^{R}_{\mathrm{cut}} = T/2$ for $\omega_1 < 0$. The choice of the cutoff time $\tau_{\mathrm{cut}}$ is varied to assess the systematic error associated with this choice.

Finally, we filter the choices of cutoff times and kinematics to demand that for a given momentum orbit and value of $\omega_1$ a minimum percentage of $\Fpigg(q_1^2, q_2^2)$ must come from the lattice data. This is done to minimize the introduction of a model dependence in the final result. To do so, we define the \emph{data content} for a given momentum and value of $\omega_1$ to be
\begin{equation}
\label{eq:data_content_def}
\Delta_{\mathrm{latt.}} = \frac{\int_{\tau^L_{\textrm{cut}}}^{\tau^R_{\textrm{cut}}} d\tau \, \tilde A^{\textrm{(latt.)}}(\tau) e^{\omega_1 \tau}}{\Fpigg(q_1^2,q_2^2)}.
\end{equation}
In this work, we always restrict kinematics to only include TFFs for which $\Delta_{\mathrm{latt.}} \geq 95\%$.

\subsection{Sampling in the Momentum Plane}
\label{sec:sampling_in_momentum_plane}
As illustrated in \Fig~\ref{fig:yield}, the transition form factor obtained from the lattice is a continuous function of $\omega_1$ for each spatial momentum orbit $|\vec{q}_1|^2$. 
In order to interpolate/extrapolate to the rest of the $(q_1^2, q_2^2)$ plane, we first sample choices of $\omega_1$ on which to evaluate the TFF as the input points for the fit.
In this work, we do so by selecting $\omega_1$ corresponding to fixed choices of the ratio $q_2^2/q_1^2$, i.e., by finding the intersection between the available orbits and several diagonal ``cuts'' through the $(q_1^2, q_2^2)$ plane.
Because the underlying lattice data are the same for all choices of $\omega_1$ within each orbit, data at nearby values of $\omega_1$ are strongly correlated. A relatively sparse sampling is therefore possible without sacrificing useful inputs to the $z$-expansion fits.
Both $q_2^2/q_1^2 = 0$ and $q_2^2/q_1^2 = 1$ are useful choices of ratios to include, the former since the single-virtual transition form factor plays an important role in the evaluation of $a_\mu^{\pi^0-\mathrm{pole}}$, and the latter since we get the best signal-to-noise ratio in the transition form factor there.

\begin{figure}[b!]
	\centering
	\includegraphics{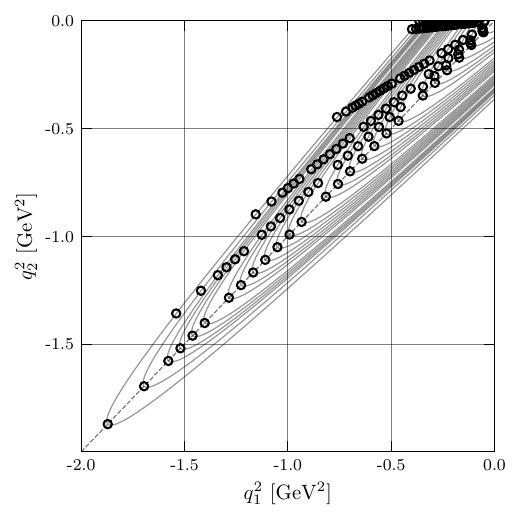}
	\caption{Cuts $(q_2^2/q_1^2) \in \{1.0, 0.88, 0.78, 0.59, 0.1,
          0.0 \}$ on ensemble cB211.072.64 illustrating our sampling
          of $\Fpigg(q_1^2, q_2^2)$ in the momentum plane used as
          input for the fit of the modified $z$-expansion. \label{fig:sampling}}
\end{figure}

We use the following procedure for the sampling in the momentum plane. 
We determine five values of 
$\omega_1$ on the highest momentum orbit $|\vec q_1|^2 = 32 (2
\pi/L)^2$, including those corresponding to the diagonal and
single-virtual kinematics,
such that the arc length of the curve
parameterized by $\omega_1$ between neighbouring samples
is constant.
This fixes five ratios $q_2^2/q_1^2$ which are then used to determine the $\omega_1$ on all other spatial momentum orbits. 
To better cover the region close to single-virtual kinematics, we additionally include a cut $q_2^2/q_1^2 = 0.1$, 
as depicted in \Fig~\ref{fig:sampling}. 
We use several subsets of these cuts
in the further analysis in order to determine systematic uncertainties
associated with the sampling: 
all combinations of three cuts
including both $q_2^2/q_1^2 \in \{1, 0\}$ and one of $q_2^2/q_1^2 \in \{0.88,
0.78, 0.59, 0.1\}$, 
and in addition a subset containing the four cuts $q_2^2/q_1^2 \in
\{1.0, 0.88, 0.1, 0.0\}$. Including more than four cuts usually leads to an
ill-conditioned covariance matrix yielding nonconverging or unstable fits. 

For each cut, we impose a threshold for a minimal data content
$\Delta_{\mathrm{latt.}} \geq 95\%$ (see
Sec.~\ref{ssec:NumericalIntegration}) in the TFF evaluated at each
sampled kinematical point, dropping those from further analysis that do not meet the threshold.

\subsection{\texorpdfstring{$z$}{z}-Expansion}
\label{ssec:z-exp}
We use the modified $z$-expansion proposed in \cite{Gerardin:2019vio} to 
interpolate/extrapolate the data from the sampling as described above to 
the whole kinematical range. 
This is a model-independent way of extending the transition form factor to arbitrary photon momenta which is preconditioned to more easily reproduce the form factor structure. 
Following Ref.~\cite{Gerardin:2019vio}, the model-independent fit form is constructed by first defining 
the conformal variables $z_1$ and $z_2$ as \cite{Boyd:1995sq}
\begin{equation}
z_k = \frac{\sqrt{t_c + Q_k^2} - \sqrt{t_c - t_0}}{\sqrt{t_c + Q_k^2} + \sqrt{t_c - t_0}}, \quad k \in \{1,2\},
\end{equation}
where $Q_k^2 = -q_k^2$ and $t_c = 4 m_{\pi}^2$ indicates the position of the branch cut due to the two-pion threshold. 
The parameter $t_0$ can be freely tuned to optimize the rate of convergence. 
For a given choice of $Q_{\mathrm{max}}^2$, to best map the region $0 \leq Q_{1,2}^2 \leq Q_{\mathrm{max}}^2$ to conformal variables near the origin, the optimal choice of $t_0$ is
\begin{equation}
t_0 = t_c \left(1 - \sqrt{1 + Q_{\mathrm{max}}^2/t_c}\right),
\end{equation}
which minimizes the maximum value of $|z_k|$ in the range $[0, Q_{\mathrm{max}}^2]$. We use $Q_{\mathrm{max}}^2 = 4$\,GeV$^2$ in the present study, as this is the region yielding the dominant contribution to the integral for $a_\mu$ (see Sec.~\ref{ssec:amu}).

In terms of these conformal variables, one can expand
\begin{equation}
\Fpigg(-Q_1^2,-Q_2^2) = \sum_{n,m=0}^\infty c_{nm} z_1^n z_2^m,
\end{equation}
which is a convergent expansion within the analytic domain $|z_k| < 1$.
The coefficients $c_{nm} = c_{mn}$ are symmetric due to the Bose symmetry. 
By restricting the sum to $m,n \leq N$, one can approximate the form factor with accuracy determined by the choice of maximum order $N$.
In addition, the transition form factor can be multiplied by an arbitrary analytic function $P(Q_1^2, Q_2^2)$ before expanding the resulting product in powers of $z_k$ as above. This preconditioning allows the expansion to more easily reproduce the form factor structure. As shown in Ref.~\cite{Gerardin:2019vio}, the choice
\begin{equation}
P(Q_1^2, Q_2^2) = 1 + \frac{Q_1^2 + Q_2^2}{M_V^2},
\end{equation}
where $M_V \approx 775\,\mathrm{MeV}$ is the $\rho$-meson mass, leads to a parameterization of the transition form factor which decreases asymptotically as $1/Q^2$ in all directions in the momentum plane even at finite values of $N$, thus satisfying the momentum dependence of \Eqs~\eqref{eq:Brodsky-Lepage} and \eqref{eq:OPE}. 
As shown in Ref.~\cite{Bourrely:2008za}, to enforce the appropriate scaling at the two-pion threshold, 
the expansion should be further modified to fix the derivatives of the transition form factors with respect to $z_k$ at $z_k = -1$ to zero. Including the cutoff $N$, preconditioning, and threshold constraints gives the modified expansion \cite{Gerardin:2019vio} 
\begin{align}
P(Q_1^2, Q_2^2) \Fpigg (-Q_1^2,&-Q_2^2) \nonumber \\ 
\approx \sum_{m,n = 0}^N c_{nm} &\left(z_1^n - (-1)^{N+n+1} \frac{n}{N+1} z_1^{N+1} \right) \nonumber \\ 
\times &\left(z_2^m - (-1)^{N+m+1} \frac{m}{N+1} z_2^{N+1} \right).\label{eq:mod_zexp_def}
\end{align}

\begin{figure*}[htbp]
	\centering
	\includegraphics{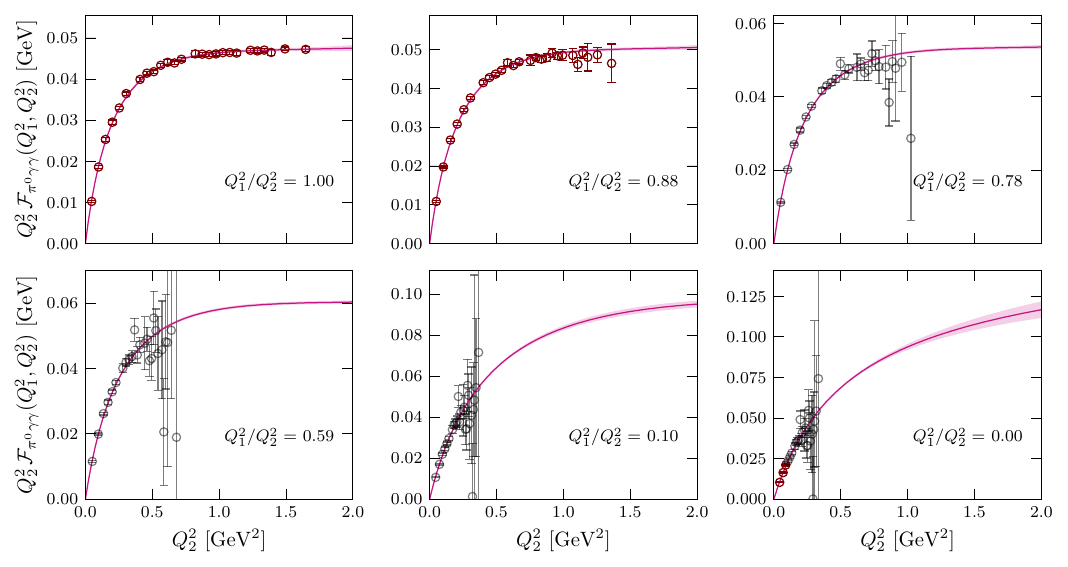}
	\caption{Correlated $z$-expansion fit to the TFFs evaluated on ensemble cC211.060.80 using $c_{00}, c_{10}, c_{11}$ and $c_{22}$. Shown is $Q_2^2 \,\Fpigg(Q_1^2,Q_2^2)$, plots showing $\Fpigg(Q_1^2,Q_2^2)$ can be found in \Fig~\ref{fig:zExpansion_TFF_only}. 
	Only TFF data points with at least 95\% data content are
        included (colored in red for the used cuts $(q_2^2 / q_1^2) \in \{1.0, 0.88, 0.0 \}$); data in grey with less than 95\% data content are shown for illustration only. For $\tilde{A}(\tau)$ a global LMD fit with fit range $[20, 21]$ and $\tau_{\mathrm{cut}} = 23$ in lattice units is used. 
	The reduced $\chi^2$ are $\chi^2_{\tilde{A}}/\mathrm{d.o.f.} = 0.86$ and $\chi^2_{z\mathrm{-exp.}}/\mathrm{d.o.f.} = 1.00$. 
         \label{fig:zExpansion}
	}
\end{figure*}
\begin{figure*}[htbp]
	\centering
	\includegraphics{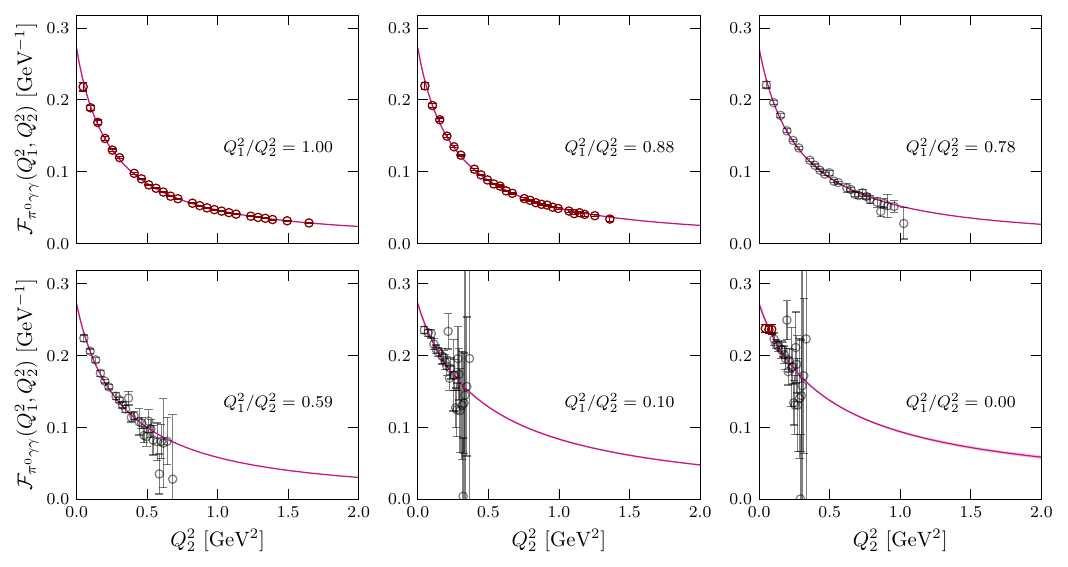}
        \caption{Same as \Fig~\ref{fig:zExpansion}, but showing
          $\Fpigg(Q_1^2,Q_2^2)$ instead of $Q_2^2 \Fpigg(Q_1^2,Q_2^2)$  in order to highlight the quality of
          fit at low values of $Q_2^2$.  \label{fig:zExpansion_TFF_only}}
\end{figure*}

In the following, we either fit the coefficients $c_{nm}$ independently to the data for each ensemble or perform a combined fit to all three ensembles with correction terms for $\mathcal{O}(a^2)$ lattice artifacts as
\begin{equation}
c_{nm}(a) = c_{nm}(0) + \delta_{nm} \left(\frac{a}{a_{\mathrm{ref.}}}\right)^2,
\end{equation}
where we use the cC211.060.80 lattice spacing for $a_{\mathrm{ref.}}$,
i.e., $a_{\mathrm{ref.}} = 0.06821(12)\,$fm. The
continuum-extrapolation strategies based on these two options are considered below in Sec.~\ref{sec:results}.

In contrast to Refs.~\cite{Gerardin:2019vio} and Ref.~\cite{Gerardin:2023naa} we perform fully correlated fits with the modified $z$-expansion. We use $N = 1,2$ and 
all possible combinations of the $N=1$ coefficients with one of the $N=2$ coefficients. 
We find that the resulting fits describe the lattice data well, even
along the cuts not included in the fit, as seen in the example in \Fig~\ref{fig:zExpansion}~and~\ref{fig:zExpansion_TFF_only}. 

Note that the transition form factors obtained from
\Eq~\eqref{eq:mod_zexp_def} multiplied by $Q^2$ asymptote to a
constant by construction, with predictions for the single-virtual and
diagonal case described in \Eqs~\eqref{eq:Brodsky-Lepage}
and~\eqref{eq:OPE}, respectively. We exclude analyses in which the
single-virtual transition form factor multiplied by $Q^2$ turn out to be asymptotically negative, since this clearly violates those predictions.

\begin{figure}[t!]
	\centering
	\includegraphics{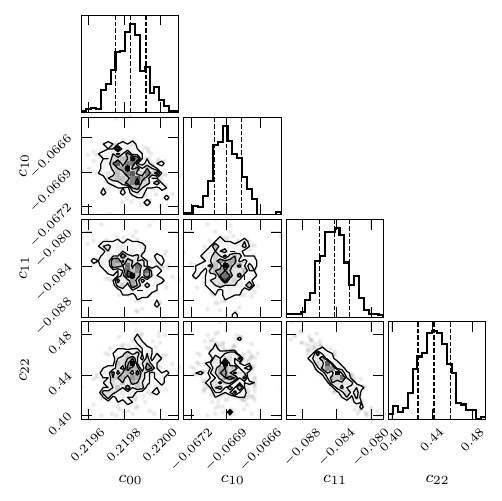}
	\caption{
	Corner plot for the coefficients $c_{00}, c_{10}, c_{11}$ and
        $c_{22}$ from the correlated $z$-expansion fit to the TFFs
        on cC211.060.80 as shown in \Fig~\ref{fig:zExpansion} and \ref{fig:zExpansion_TFF_only} from a jackknife resampling.
        The corresponding correlation matrix is given in \Eq~\eqref{eq:ex_corrmat}.
	\label{fig:zExpansionCoefficients}}
\end{figure}

We find that correlated $z$-expansion fits with cutoffs set to either $N = 1$ or $N = 2$ already lead to 
a good quality of fit 
when including input data from three or four cuts in the momentum
plane as described in Sec.~\ref{sec:sampling_in_momentum_plane}. 
For $N = 2$, at least two of the coefficients $c_{nm}$ are strongly
correlated, indicating a degeneracy in the parameterization. Even when
including only one of the second-order coefficients in conjunction
with all three first-order coefficients, we find slight correlations
between some of them. See \Fig~\ref{fig:zExpansionCoefficients} for a
corner plot depicting correlations in a representative case when
including coefficients $\{ c_{00}, c_{10}, c_{11}, c_{22} \}$.

The corresponding correlation matrix is given by
\begin{equation}
\label{eq:ex_corrmat}
\mathrm{cor}(c_{nm}) = \left( \begin{array}{cccc} 
				+1.00 & -0.30 & -0.39 & +0.31 \\
				-0.30 & +1.00 & -0.05 & -0.14 \\
				-0.39 & -0.05 & +1.00 & -0.88 \\
				+0.31 & -0.14 & -0.88 & +1.00 
			\end{array} \right),
\end{equation}
clearly showing the strong (anti)correlation between $c_{11}$ and
$c_{22}$, and lesser (anti)correlations between all other coefficients except for $c_{10}$ and $c_{11}$ which are almost completely uncorrelated. When including all $N=2$ coefficients, it turns out that $c_{20}$ and  $c_{21}$ are strongly correlated with $c_{11}$ for almost all choices of parameters.
To get reliable estimates of the individual coefficients, it is desirable to remove such near-degeneracies in the parameterization by only using a subset of the coefficients which are not strongly correlated. 
In the final analysis, we therefore determine the preferred values of
all quantities studied based on a $z$-expansion parameterization with
minimal degeneracy, but we use variations between these choices of
parameterizations to determine an additional systematic error. 

\section{Observables}
\label{sec:observables}
The main observable studied in this work is $\amupi$, i.e., the pion-pole contribution to the muon $(g-2)/2$. 
We further study the two-photon decay width $\Gamma(\pi \to \gamma
\gamma)$ and the slope parameter $b_\pi$, both of which are related to $\Fpigg$. 

\subsection{Pion-pole contribution to \texorpdfstring{$a_\mu$}{a\_mu}}
\label{ssec:amu}
\begin{figure*}[t!]
	\centering
	\includegraphics{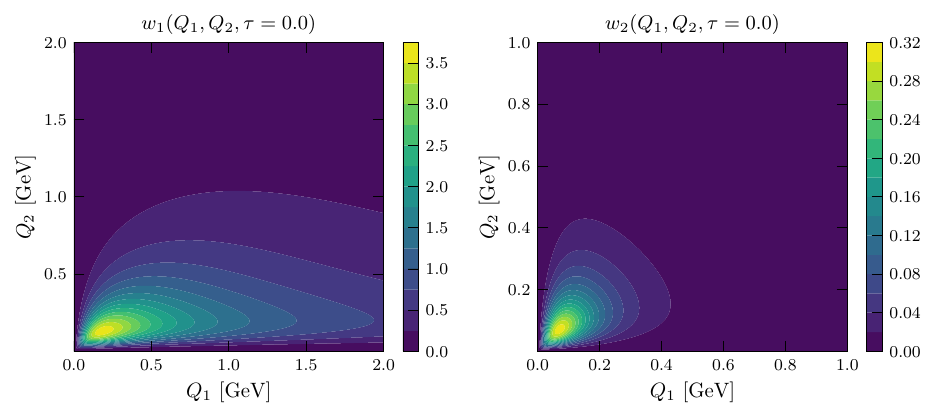}
	\caption{Weight function $w_1(Q_1, Q_2, \tau)$ (left) and $w_2(Q_1, Q_2, \tau)$ (right) for the pion for momenta $Q_1$ and $Q_2$ 
	and $\tau = \cos \theta$ where $\theta = 90^\circ$, 
	showing the concentration at low momenta. 
	Note the different momentum range for and magnitude of $w_2$ as well as its symmetry with respect to $Q_1 \leftrightarrow Q_2$. 
	\label{fig:int_rep_weights}}
\end{figure*}

Following Refs.~\cite{PhysRevD.94.053006}~and~\cite{Jegerlehner2009}, we use the three-dimensional integral representation for the pion-pole contribution
\begin{align}
\amupi &= \left(\frac{\alpha}{\pi}\right)^3 \left[ a_\mu^{\pi^0-\mathrm{pole}(1)} + a_\mu^{\pi^0-\mathrm{pole}(2)} \right],
\end{align}
where $\alpha$ is the fine structure constant,
\begin{equation}
\begin{aligned}
a_\mu^{\pi^0-\mathrm{pole}(1)} &= \int_0^\infty dQ_1 \int_0^\infty dQ_2 \int_{-1}^1 d\tau \, \\ & w_1(Q_1, Q_2, \tau) \Fpigg(-Q_1^2, -Q_3^2) \Fpigg(-Q_2^2,0), \label{eq:int_rep_1}
\end{aligned}
\end{equation}
and
\begin{equation}
\begin{aligned}
a_\mu^{\pi^0-\mathrm{pole}(2)} &= \int_0^\infty dQ_1 \int_0^\infty dQ_2 \int_{-1}^1 d\tau \, \\ & w_2(Q_1, Q_2, \tau) \Fpigg(-Q_1^2, -Q_2^2) \Fpigg(-Q_3^2,0). \label{eq:int_rep_2}
\end{aligned}
\end{equation}
The integration is performed over the magnitudes $Q_1$, $Q_2$ of two of the four-momenta and $\tau = \cos \theta$ describing the angle $\theta$ between them,
with the magnitude of the third four-momentum fixed by $Q_3^2 = Q_1^2 + Q_2^2 + 2 Q_1 Q_2 \tau$. 
The weight functions $w_1$ and $w_2$ are given
in \app{appendix:3d_int_rep}. Note that $w_{1,2}$ are both dimensionless and $w_{1,2}(Q_1, Q_2, \tau) \rightarrow 0$ for $Q_{1,2} \rightarrow 0$ and for $\tau \rightarrow \pm 1$. 
Further, $w_2$ is symmetric under the exchange of $Q_1$ and $Q_2$.
In Ref.~\cite{PhysRevD.94.053006}, the weight functions for the pion are studied and discussed in detail. One finds that the momentum region $Q_{1,2} \leq 0.5\,$GeV is the most important 
in \Eqs~\eqref{eq:int_rep_1} and \eqref{eq:int_rep_2} for $\amupi$, 
which is where we have the strongest data support from our lattice calculation. 
For two examples illustrating the concentration of the weight functions at low momenta, see \Fig~\ref{fig:int_rep_weights}. 
In particular, note that $w_2(Q_1, Q_2, \tau)$ is roughly an order of magnitude smaller than $w_1(Q_1, Q_2, \tau)$.

\begin{figure}
	\centering
	\includegraphics{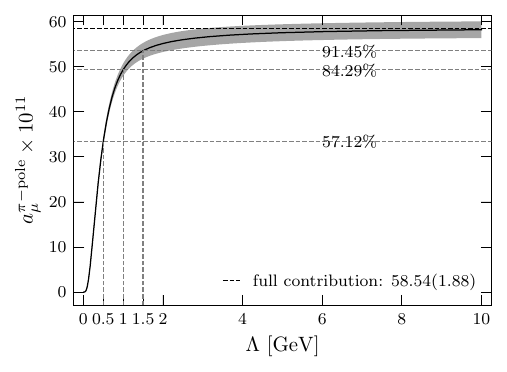}
	\caption{Contribution to $\amupi$ for increasing cutoff $\Lambda$ for ensemble cC211.060.80,
	using a $N=2$ correlated $z$-expansion fit to $(q_2^2 / q_1^2) \in \{1.0, 0.88, 0.1, 0.0 \}$.
	Only TFF data points where at least 95\% of the transition form factors come from lattice data are included, for $\tilde{A}(\tau)$ a global LMD fit with fit range $[7,8]$ and $\tau_{\mathrm{cut}} = 26$ in lattice units is used. 
	The reduced $\chi^2$ are $\chi^2_{\tilde{A}}/\mathrm{d.o.f.} = 0.85$ and $\chi^2_{z~\mathrm{exp.}}/\mathrm{d.o.f.} = 1.02$. 
	Indicated is the saturation at $\Lambda \in \{0.5, 1.0, 1.5 \}$\,GeV as well as the value of the full contribution for this particular choice of parameters. 
	\label{fig:3d_int_rep_saturation}}
\end{figure}

By introducing a momentum cutoff $\Lambda$, i.e., by replacing
\begin{equation}
    \int_0^{\infty} dQ_{1,2} \; \longrightarrow \;
    \int_0^{\Lambda} dQ_{1,2}
\end{equation}
in \Eqs~\eqref{eq:int_rep_1}~and~\eqref{eq:int_rep_2}, 
one can estimate the importance of various momentum regions in the evaluation of \Eqs~\eqref{eq:int_rep_1}~and~\eqref{eq:int_rep_2}. 
Though the cutoffs are only directly imposed on $Q_{1,2}$, they imply a cutoff of $Q_3 \leq 2 \Lambda$ on the third momentum as well. 
We find that the integrals saturate rapidly, such that we get around 85\% of the total result with $\Lambda = 1$\,GeV and around 90\% with $\Lambda = 1.5$\,GeV, i.e., from momentum regions with strong data support from our lattice calculation. 
An example of this exercise is shown in \Fig~\ref{fig:3d_int_rep_saturation}. 

\subsection{Decay Width and Slope Parameter}
\label{ssec:gamma_bP}

To leading order in the fine structure constant $\alpha$, the transition form factors determine the partial decay width through
\begin{equation}
\label{eq:decay_width}
\Gamma(\pi \to \gamma \gamma) = \frac{\pi \alpha^2 m_\pi^3}{4} \left| \Fpigg(0,0) \right|^2 + O(\alpha^3).
\end{equation}
The neutral-pion decay width has been measured in the PrimEx and PrimEx-II experiments \cite{PrimEx:2010fvg, doi:10.1126/science.aay6641} with a combined result of $\Gamma(\pi \to \gamma \gamma) = 7.802(52)_{\mathrm{stat}}(105)_{\mathrm{sys}}(117)_{\mathrm{tot}}$\,eV. At our level of statistical precision, uncertainties from radiative corrections to the decay width are negligible and we therefore quote the leading-order result as our estimate of this quantity.

Further, the transition form factors can be used to extract the slope parameter
\begin{equation}
\label{eq:slope_parameter}
b_\pi = \frac{1}{\Fpigg(0,0)} \frac{\mathrm{d} \Fpigg(q^2,0)}{\mathrm{d}q^2} \Bigg|_{q^2=0},
\end{equation}
thus providing input for determining the electromagnetic interaction radius of the pion. 
The averaged experimental result for the slope parameter is $b_\pi = 1.84(17)$\,GeV$^{-2}$ \cite{Workman:2022ynf}. 

These quantities are easily extracted from the form of the $z$-expansion fit at $(q_1^2, q_2^2) =  (0,0)$. The value of $\Fpigg(0,0)$ comes directly from the $z$-expansion fit, while the derivative can be acquired by differentiating the form of the $z$-expansion in \Eq~\eqref{eq:mod_zexp_def} with respect to $-Q_1^2$, yielding the slope parameter
\begin{equation}
\begin{aligned}
    b_{\pi} &= -\frac{d}{dQ_1^2} \ln \Fpigg(-Q_1^2,0)\big|_{Q_1^2=0} \\
    &\approx  \frac{1}{M_V^2} - \frac{1}{z_1} \frac{dz_1}{dQ_1^2}
    \Big[ \sum_{\substack{m = 0 \\ n = 1}}^N n c_{nm} \left( z_1^{n} - (-1)^{N+n+1} z_1^{N+1} \right) \\
    &\hspace{55pt} \times \left( z_2^m - (-1)^{N+m+1} \frac{m}{N+1} z_2^{N+1} \right)\Big]_{Q_{1,2}^2=0} .
\end{aligned}
\end{equation}
The term in square brackets can be directly evaluated, and the derivative of the conformal parameter is given by
\begin{equation}
    \frac{1}{z_1} \frac{d z_1}{d Q_1^2}\Bigg|_{Q_1^2 = 0} = \frac{1}{t_0} \sqrt{1 - t_0/t_c}.
\end{equation}

\section{Model averaging and error estimation}
\label{sec:model_averaging}

We follow different analysis chains from the initial lattice correlation function data to
estimates of the quantities $\amupi$, $\Gamma(\pi \to \gamma \gamma)$ and $b_\pi$. Each such chain is
determined by a choice of parameters, namely 
the choice of fit range and fit model in the fit to $\tilde{A}(\tau)$, 
$\tau_{\mathrm{cut}}$ when constructing the transition form factors and finally the included cuts in the momentum plane in the fit to the modified $z$-expansion. 
This yields $\mathcal{O}(10^3-10^4)$ estimates.
For the remainder of this section such a choice of parameters is called an ``analysis''. A priori, only fits with $\chi^2_{\tilde{A}}/\mathrm{d.o.f.}$ close to 1
are included in an analysis chain for the single ensemble analyses, with $0.84 \lesssim \chi^2_{\tilde{A}}/\mathrm{d.o.f.} \lesssim 2.05$. 
For the combined $z$-expansion fits the input $\tilde{A}(\tau)$ fit ranges and $\tau_{\mathrm{cut}}$ were chosen to be similar in physical units across all ensembles, using $t_{min} \in [0.48, 1.27]$\,fm, $t_{max} \in [0.64, 1.43]$\,fm and $\tau_{\mathrm{cut}} \in [1.35, 2.07]$\,fm on cB211.072.64 as a reference and remaining within $10\%$ of these physical-units values for the cC211.060.80 and cD211.054.96 ensembles.
The fits across all ensembles are of good quality, with $\chi^2$ values in the range $1 \lesssim \chi^2_{\tilde{A}}/\mathrm{d.o.f.} \lesssim 2.4$ with an average of $\bar{\chi}^2_{\tilde{A}}/\mathrm{d.o.f.} \approx 1.5$.

We calculate the statistical errors for each analysis using the jackknife resampling procedure, finding that there is virtually no autocorrelation between the input data $\tilde{A}(\tau)$ amongst the available configurations listed in \Tab~\ref{table:ensemble_params}.
Jackknife resampling is applied over the entire analysis pipeline for each set of analysis choices, ensuring that statistical errors are fully propagated.

\begin{figure}
	\centering
	\includegraphics{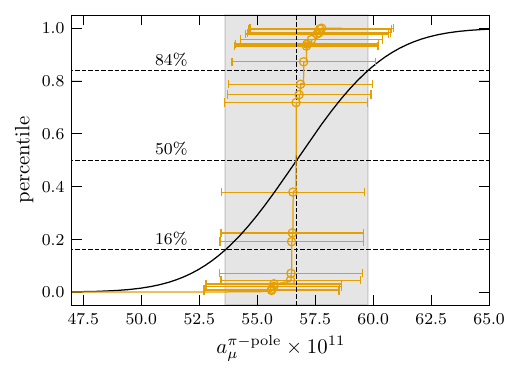}
	\caption{Cumulative distribution functions (CDF) of $\amupi$ for the combined fit using $c_{nm} = \{ c_{00}, c_{10}, c_{11}, c_{22} \}$. The orange curve shows the CDF of $\approx$15'000 different analyses obtained from their AIC weights, the errors on the orange point show the statistical errors for some analyses with a significant weight in the model averaging. This curve corresponds to statistical errors rescaled by $\lambda=0$ while the black curve corresponds to rescaling by $\lambda=1$. The latter gives the central value by the $50\%$ percentile, and the total error as the half variation between the $16\%$ and $84\%$ percentiles shown in gray.
    To separate the statistical and systematic part of the error, we further calculate the error of the distribution with statistical errors rescaled by $\lambda = 2$ (not shown in the figure), as described in the main text. 
	\label{fig:CDF}}
\end{figure}

To take a weighted average of estimates under various analysis choices in an analysis chain, 
we use a modified version of the Akaike Information Criterion (AIC)~\cite{Akaike1978,Jay2021}. 
Closely following the method introduced in~\cite{doi:10.1126/science.1257050} and~\cite{Borsanyi:2020mff}, the model averaging proceeds as follows. For a target observable $y$, here $\amupi$, $\Gamma(\pi \to \gamma \gamma)$ or $b_\pi$, we build a histogram from the different analyses, assigning to each analysis a weight given by the AIC. This criterion is derived from the Kullback-Leibler divergence, which measures the distance of a fit function from the true distribution of the points (for a derivation see Ref.~\cite{doi:10.1126/science.1257050}). 
We use the modified AIC introduced in~\cite{Borsanyi:2020mff},
\begin{equation}
\label{eq:AIC}
\text{AIC} \sim \exp\left[ -\frac{1}{2}\left( \chi^2 + 2 n_{\text{par}} - n_{\text{data}} \right) \right],
\end{equation}
where the $\chi^2$, the number of fit parameters $n_{\text{par}}$ and the number of data points $n_{\text{data}}$ describe the fit of interest. 
The first two terms in the exponent correspond to the standard AIC, and the last term is needed to weigh fits with different lengths in the fit ranges when fitting to $\tilde A (\tau)$ or to weight $z$-expansion fits with a differing number of cuts in the momentum plane when sampling the transition form factors. 

In general, two fitting steps are involved in the determination of each of $\amupi$, $\Gamma(\pi^0 \to \gamma\gamma)$, and $b_\pi$ on a given ensemble: the global fit to $\tilde A(\tau)$ and the $z$-expansion fit to the transition form factors. A combined weight can be assigned to each combination of fitting choices across both of these steps by multiplying the respective AIC weights $\tilde{w}_i$ given in \Eq~\eqref{eq:AIC}. In the alternative simultaneous continuum and $z$-expansion fit procedure described in Sec.~\ref{ssec:z-exp}, the model averaging step is instead performed globally across all three ensembles, and therefore a combined AIC weight is assigned by multiplying the weights of the fits to $\tilde{A}(\tau)$ across all three ensembles as well as the weight of the final global $z$-expansion fit with lattice artifacts included. In either case, the result is one unnormalized weight $\tilde{w}_i$ per analysis.

Given the weights $\tilde{w}_i$, central values $m_i$, and statistical error $\sigma_i$ from each analysis, the global cumulative distribution function (CDF) is expected to be well-described by a weighted combination of normal distribution CDFs. We take the median of the CDF as the central value across all analyses and the half variation between the 16\% and 84\% percentiles as the total error. To separate statistical and systematic errors, the same procedure can be performed with errors rescaled as $\sigma_i \rightarrow \sqrt{\lambda} \sigma_i$. The variation between $\lambda = 2$ and $\lambda = 1$ yields a separate statistical and systematic error~\cite{Borsanyi:2020mff}.
For an illustration, see \Fig~\ref{fig:CDF}. 

For a better understanding on the composition of the systematic error, we follow the error budgeting procedure suggested in \cite{Borsanyi:2020mff}: For each of the choices made during the analysis chain, e.g., the choice between the VMD or LMD fit to $\tilde A(\tau)$ or the different fit ranges, we first determine the total error for each possible option, varying all other components of the analysis. We then construct a second CDF as above
with $m_i$ the average of the 16\% and 84\% percentiles, $\sigma_i$ the total error and $w_i$ the sum of the weights coming from this choice. Using this CDF, we derive the systematic error as described above for the original CDF; this is our result for the systematic error corresponding to the choice. Note that the estimated systematic errors associated with each of the steps of the analysis by this procedure
are correlated, thus they do not sum up quadratically to the full systematic error.

\section{Lattice results and continuum extrapolation}
\label{sec:results}
Here, our results using the model averaging described in Sec.~\ref{sec:model_averaging} are summarized. Comparing to our earlier publication \cite{BurriPoS2022} we have refined the analysis by systematically studying different $z$-expansion fits and by excluding analyses leading to unphysical TFFs as described in Sec.~\ref{ssec:z-exp}. 

\begin{figure*}[ht!]
	\centering
	\includegraphics{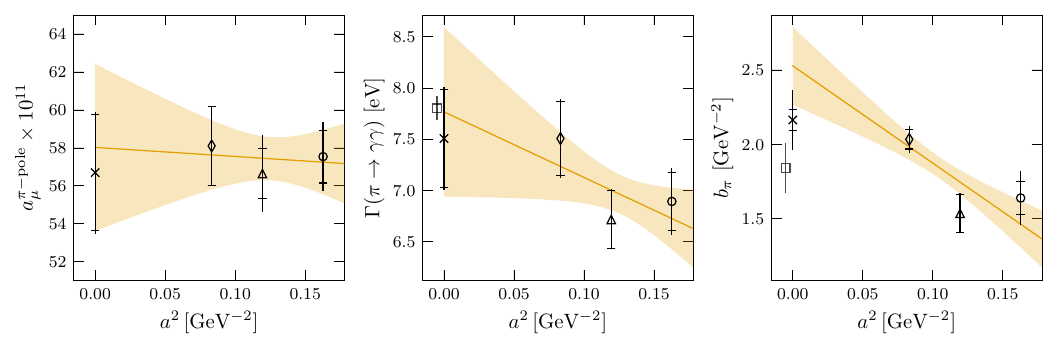}
	\caption{Combined fit and single ensemble results from \Tabs~\ref{table:single_ensemble_results}~and~\ref{table:combined_fit_results}: $\amupi$ (left), $\Gamma(\pi \to \gamma \gamma)$ (middle) and $b_\pi$ (right). Indicated are the statistical and total errors. 
	For comparison, we also show linear fits in $a^2$ on the per-ensemble data points. 
	In each plot, the points with caps on the errorbars
        correspond to the combined fit (cross), the cD211.054.96 (diamond), the cC211.060.80 (triangle) and the cB211.072.64 (circle) result. 
	In the $\Gamma(\pi \to \gamma \gamma)$ and $b_\pi$ plots, the
        square symbols
        show the experimental values and their errors, namely $\Gamma(\pi \to \gamma \gamma) = 7.802(117)$\,eV \cite{doi:10.1126/science.aay6641} and $b_\pi = 1.84(17)$\,GeV$^{-2}$ \cite{Workman:2022ynf}.
	\label{fig:results}}
\end{figure*}

For the single ensemble analyses, we perform an AIC averaging over
different fit models and fit ranges for $\tilde A(\tau)$, different
choices of $\tau_{\mathrm{cut}} \approx [1.3, 2.0]$\,fm on all three
ensembles, different samplings in the momentum plane and different choices of $z$-expansion fits. 
The results including error budgeting are summarized in
\Tab~\ref{table:single_ensemble_results}, and are shown in
\Fig~\ref{fig:results} as diamonds (cD211.054.96), triangles
(cC211.060.80), and circles (cB211.072.64), respectively. 	We note that the error
budgets from fit model and fit range are very small, indicating that
by restricting the transition form factors in the $z$-expansion to
have at least a 95\% contribution from lattice data effectively removes the
dependence on the model used to fit the tails of $\tilde{A}(\tau)$ almost
completely.
	\begin{table}
	\centering
	\begin{tabular}{||c | c c c||} 
		\hline
		cB211.072.64 & $\amupi \times 10^{11}$ & $\Gamma(\pi \to \gamma \gamma)$\,[eV] & $b_\pi$\,[GeV$^{-2}$] \\ 
		\hline
		value & 57.5(1.8) & 6.89(0.33) & 1.63(0.18)  \\ 
		$\sigma_{\mathrm{stat}}$ & 1.39 & 0.28 & 0.11  \\ 
		$\sigma_{\mathrm{sys}}$ & 1.16 & 0.18 & 0.15  \\ 
		\hline \hline
		fit model & 0.09 & 0.01 & 0.01 \\
		fit range & 0.00 & 0.00 & 0.00 \\
		$\tau_{\mathrm{cut}}$ & 0.17 & 0.06 & 0.03 \\
		sampling & 0.20 & 0.06 & 0.04 \\
		$z$-exp. & 1.15 & 0.17 & 0.14 \\
		\hline
	\end{tabular}
	\begin{tabular}{||c | c c c||} 
		\hline
		cC211.060.80 & $\amupi \times 10^{11}$ & $\Gamma(\pi \to \gamma \gamma)$\,[eV] & $b_\pi$\,[GeV$^{-2}$] \\ 
		\hline
		value & 56.7(2.0) & 6.72(0.29) & 1.54(0.14)  \\ 
		$\sigma_{\mathrm{stat}}$ & 1.33 & 0.28 & 0.13  \\ 
		$\sigma_{\mathrm{sys}}$ & 1.54 & 0.08 & 0.04  \\ 
		\hline \hline
		fit model & 0.01 & 0.00 & 0.00 \\
		fit range & 0.07 & 0.01 & 0.00 \\
		$\tau_{\mathrm{cut}}$ & 0.36 & 0.03 & 0.02 \\
		sampling & 0.50 & 0.03 & 0.03 \\
		$z$-exp. & 1.51 & 0.04 & 0.02 \\
		\hline
	\end{tabular}
	\begin{tabular}{||c | c c c||} 
		\hline
		cD211.054.96 & $\amupi \times 10^{11}$ & $\Gamma(\pi \to \gamma \gamma)$\,[eV] & $b_\pi$\,[GeV$^{-2}$] \\ 
		\hline
		value & 58.1(2.1) & 7.51(0.39) & 2.03(0.09)  \\ 
		$\sigma_{\mathrm{stat}}$ & 2.08 & 0.36 & 0.06  \\ 
		$\sigma_{\mathrm{sys}}$ & 0.37 & 0.13 & 0.06  \\ 
		\hline \hline
		fit model & 0.00 & 0.00 & 0.00 \\
		fit range & 0.01 & 0.00 & 0.00 \\
		$\tau_{\mathrm{cut}}$ & 0.19 & 0.08 & 0.03 \\
		sampling & 0.10 & 0.05 & 0.03 \\
		$z$-exp. & 0.29 & 0.10 & 0.06 \\
		\hline
	\end{tabular}
	\caption{Values of measured observables on each ensemble, with systematic error budgeting indicated below each result.
	\label{table:single_ensemble_results}}
	\end{table}

\begin{figure}[ht!]
	\centering
	\includegraphics{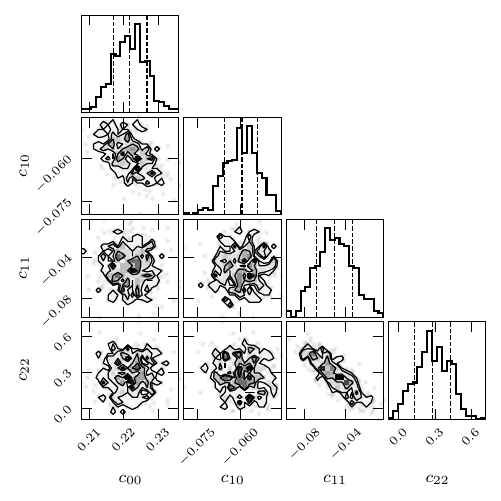}
	\caption{
	Corner plot of the AIC averaged coefficients of the combined correlated $z$-expansion fit, utilizing bootstrap resampling on each ensemble, where only the coefficients $c_{00}, c_{10}, c_{11}$ and $c_{22}$ were used. 
	Notice the anticorrelation between $c_{11}$ and $c_{22}$ and the less severe anticorrelation between  $c_{00}$ and $c_{10}$. All other coefficients are virtually uncorrelated. The corresponding correlation matrix is given in \Eq~\eqref{eq:cont_corrmat}. 
	\label{fig:zExpansionCoefficients_AIC_avg}}
    \end{figure}

\begin{figure*}[ht!]
	\centering
	\includegraphics{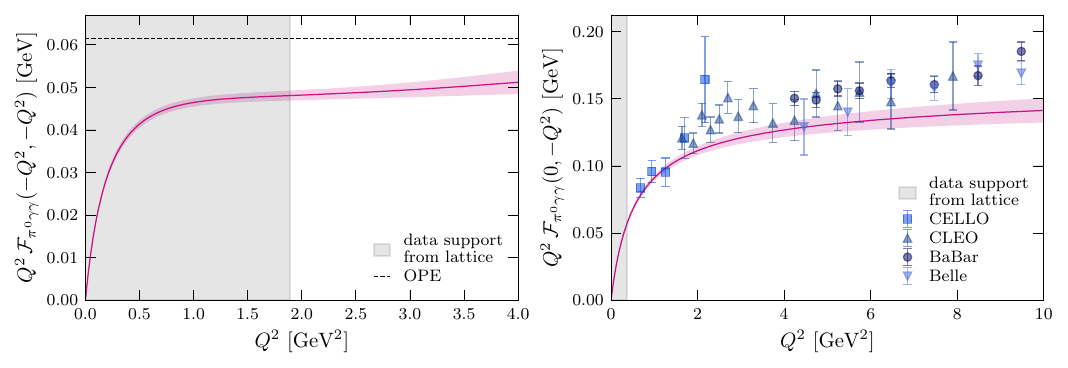}
	\caption{Transition form factor from the combined fit for
          diagonal (left) and single-virtual (right) kinematics. For
          the diagonal kinematics, the OPE prediction for large $Q^2$ is
          indicated by the horizontal dashed line while for the single-virtual kinematics, experimental values from CELLO \cite{CELLO:1990klc}, CLEO \cite{CLEO:1997fho}, BaBar \cite{BaBar:2009rrj, BaBar:2011nrp} and Belle \cite{Belle:Uehara2012} are shown. The region with direct support from lattice data is shaded in grey. 
	\label{fig:comb_fit_TFFs}}
\end{figure*}

For the $z$-expansion fits considered in the combined fitting, the set
$c_{nm} = \{ c_{00}, c_{10}, c_{11}, c_{22} \}$ with the
lattice-artefact correction coefficients $\delta_{nm} =
\{\delta_{00}, \delta_{10}\}$ leads to the least
correlation amongst the coefficients and gives an AIC-averaged fully
correlated $\chi^2/\mathrm{d.o.f.}=1.36$.
The AIC-averaged $z$-expansion coefficients
(cf.~\Eq~\eqref{eq:mod_zexp_def}) in the continuum limit based on the combined fit across all ensembles are given by
\begin{center}
\begin{tabular}{||c c c c||} 
		\hline
		$c_{00}$ & $c_{10}$ & $c_{11}$ & $c_{22}$ \\
		\hline
		0.2220(48) & -0.0596(59) & -0.050(18) & 0.27(14) \\
		\hline
\end{tabular}
\end{center}
with correlation matrix
\begin{equation}
\label{eq:cont_corrmat}
\mathrm{cor}(c_{nm}) = \left( \begin{array}{cccc} 
				+1.00 & -0.46 & -0.07 & +0.07 \\
				-0.46 & +1.00 & +0.03 & -0.08 \\
				-0.07 & +0.03 & +1.00 & -0.83 \\
				+0.07 & -0.08 & -0.83 & +1.00 
			\end{array} \right)
\end{equation}
and the corresponding corner plot given in \Fig~\ref{fig:zExpansionCoefficients_AIC_avg}. 
This is our preferred continuum result. We note that the coefficients
describing the lattice artefacts are very small and compatible with
zero, $\delta_{00}=0.0005(41)$ and $\delta_{10}=-0.0085(49)$.
The resulting continuum TFFs for diagonal and single-virtual
kinematics are shown in \Fig~\ref{fig:comb_fit_TFFs}. The figure
further indicates the kinematic region directly supported by lattice
data, where the fit is expected to be most reliable. For the
single-virtual kinematics we find good
agreement with the CELLO \cite{CELLO:1990klc} bins
below $Q^2 \lesssim
2\text{ GeV}^2$, while at larger momenta $Q^2 \gtrsim 2\text{
  GeV}^2$ our analysis results tend to be systematically
lower than the experimental ones. 

For the quantities $\amupi$, $\Gamma(\pi \to \gamma \gamma)$ and
$b_\pi$, in order to account for different choices of $z$-expansion parameters in a more
conservative way than by using the AIC, we use a procedure inspired by
the
method described in
Ref.~\cite{ExtendedTwistedMass:2021gbo}. It consists of taking the
absolute difference of the central value of the combined fit using the
coefficient set $c_{nm} = \{ c_{00}, c_{10}, c_{11}, c_{22} \}$ and
the weighted average of the results from the other considered fits each weighted with $1/\sigma_{\mathrm{tot}}^2$ as an additional systematic error $\sigma_{\mathrm{sys,}\,z\mathrm{-exp.}}$  added in quadrature. 
We find
\begin{equation}
  \begin{gathered}
\amupi = 56.7(3.1)_{\mathrm{stat}}(1.0)_{\mathrm{sys}}[3.2]_{\mathrm{tot}} \times 10^{-11}, \\
\Gamma(\pi \to \gamma \gamma) = 7.50(0.48)_{\mathrm{stat}}(0.16)_{\mathrm{sys}}[0.50]_{\mathrm{tot}}\,\mathrm{eV} ,\\
b_\pi =
2.16(0.07)_{\mathrm{stat}}(0.19)_{\mathrm{sys}}[0.20]_{\mathrm{tot}}\,\mathrm{GeV}^{-2}.
\label{eq:lattice results}
\end{gathered}
\end{equation}
These values 
including
a detailed error budget are summarized in
\Tab~\ref{table:combined_fit_results}, 
	\begin{table}
	\centering
	\begin{tabular}{||c | c c c||} 
		\hline
		combined fit & $\amupi \times 10^{11}$ & $\Gamma(\pi \to \gamma \gamma)$\,[eV] & $b_\pi$\,[GeV$^{-2}$] \\ 
		\hline
		value & 56.7(3.2) & 7.50(0.50) & 2.16(0.20)  \\ 
		$\sigma_{\mathrm{stat}}$ & 3.06 & 0.48 & 0.07  \\ 
		$\sigma_{\mathrm{sys,}\,z\mathrm{-exp.}}$ & 0.91 & 0.14 & 0.19  \\ 
		$\sigma_{\mathrm{sys}}$ & 0.35 & 0.07 & 0.01  \\ 
		\hline \hline
		fit model & 0.00 & 0.00 & 0.00 \\
		fit range & 0.00 & 0.00 & 0.00 \\
		$\tau_{\mathrm{cut}}$ & 0.33 & 0.07 & 0.00 \\
		sampling & 0.19 & 0.05 & 0.01 \\
		\hline 
	\end{tabular}
	\caption{Preferred continuum result from the combined fit
          using $c_{nm} = \{ c_{00}, c_{10}, c_{11}, c_{22} \}$ and
          $\delta_{nm} = \{ \delta_{00}, \delta_{10}
          \}$. $\sigma_{\mathrm{sys,}\,z\mathrm{-exp.}}$ denotes the
          more conservative systematic error from different $z$-expansion
          parameter choices as described in the text.
	\label{table:combined_fit_results}}
	\end{table}
and are shown in
\Fig~\ref{fig:results} as crosses.
Instead of using the continuum values of the $z$-expansion coefficients from the combined fit to calculate the continuum values of $\amupi$, $\Gamma(\pi \to \gamma \gamma)$ and $b_\pi$, 
we could also extrapolate their per-ensemble values in $a^2$
directly. As depicted in \Fig~\ref{fig:results}, we find that fits
linear in $a^2$ agree well with the combined fit continuum results,
albeit with significantly larger errors at $a^2 = 0$. We note that at
the current level of accuracy the
lattice artefacts in $\amupi$ are compatible with zero. Hence, a
continuum extrapolation with a constant is also possible leading to  a
better $\chi^2/\mathrm{d.o.f.}$ and a
significantly smaller error. However, in order to be on the
conservative side, we do not consider this value in our analysis.  

Our result for $\amupi$ is compatible with our earlier
analysis \cite{BurriPoS2022}, where we used a continuum extrapolation
in the individual ensemble estimates of each observable instead of one
coming from a combined fit to the TFF across ensembles. It is also
compatible with the recent lattice result $\amupi =
57.8\pm1.8_{\mathrm{stat}}\pm0.9_{\mathrm{sys}} \times 10^{-11}$ from
the BMW collaboration in
Ref.~\cite{Gerardin:2023naa}, and $\amupi = 59.7\pm3.6  \times 10^{-11}$ from the Mainz group in Ref.~\cite{Gerardin:2019vio}. 
Comparing to the dispersive result $\amupi = 63.0^{+2.7}_{-2.1} \cdot 10^{-11}$ from Refs.~\cite{Aoyama:2020ynm,Hoferichter:2018dmo,Hoferichter:2018kwz}, we find that our result is compatible at the level of 1.6\,$\sigma$.

For $\Gamma(\pi \to \gamma \gamma)$, we again find agreement with $\Gamma(\pi \to \gamma \gamma) = 7.11\pm0.44_{\mathrm{stat}}\pm0.21_{\mathrm{sys}}$\,GeV$^{-2}$ from Ref.~\cite{Gerardin:2023naa}. Our result is also compatible with the experimental value $\Gamma(\pi \to \gamma \gamma) = 7.802(52)_{\mathrm{stat}}(105)_{\mathrm{sys}}$\,eV from Ref.~\cite{doi:10.1126/science.aay6641}.

Finally, for $b_\pi$, our result is compatible with the experimental result $b_\pi = 1.84(17)$\,GeV$^{-2}$ from Ref.~\cite{Workman:2022ynf} at the level of 1.2\,$\sigma$. 
We also agree at the level of 1.6\,$\sigma$ with $b_\pi = 1.78(12)$\,GeV$^{-2}$ from the extraction based on Pad\'e approximants \cite{Masjuan:2012wy} and find a slight tension of 2.1\,$\sigma$ with the dispersive result $b_\pi = 1.73(5)$\,GeV$^{-2}$ from Refs.~\cite{Hoferichter:2018dmo, Hoferichter:2018kwz}. In all cases our central value is higher than these results by $O(10\%)$. Further investigation of this quantity from the lattice may therefore be of interest.

\section{Combined analysis of lattice and experimental data}
\label{sec:lattice+exp analysis}
With our determination of the TFF at the physical point and in the
continuum limit we can attempt a combined analysis of our lattice data
together with the available experimental data from CELLO \cite{CELLO:1990klc},
CLEO \cite{CLEO:1997fho}, BaBar \cite{BaBar:2009rrj, BaBar:2011nrp}
and Belle \cite{Belle:Uehara2012}. Such an exercise is interesting for
several reasons.

\begin{figure*}[htb!]
	\centering
	\includegraphics{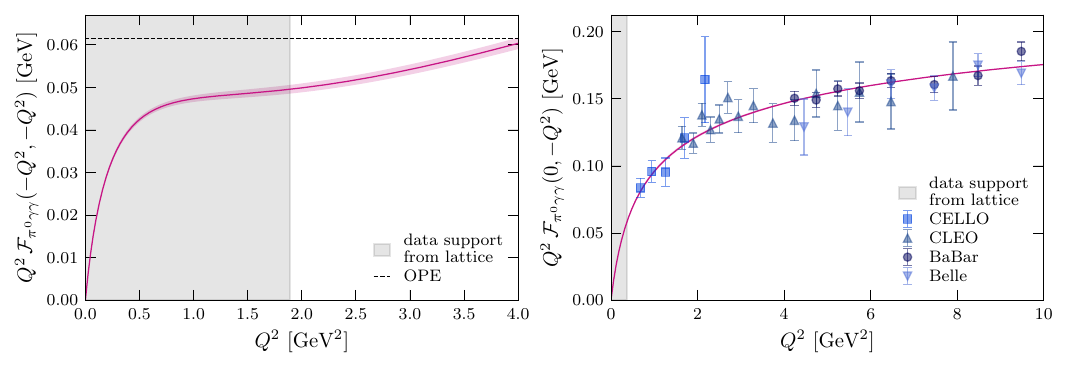}
	\caption{Transition form factor from the combined fit to
          lattice and experimental data for
          diagonal (left) and single-virtual (right) kinematics. For
          the diagonal kinematics, the OPE prediction for large $Q^2$ is
          indicated by the horizontal dashed line while for the single-virtual kinematics, experimental values from CELLO \cite{CELLO:1990klc}, CLEO \cite{CLEO:1997fho}, BaBar \cite{BaBar:2009rrj, BaBar:2011nrp} and Belle \cite{Belle:Uehara2012} are shown. The region with direct support from lattice data is shaded in grey. 
	\label{fig:comb_latt+exp_fit_TFFs}}
\end{figure*}

Firstly, we can test whether or not the systematic
difference, as evident from
\Fig~\ref{fig:comb_fit_TFFs}, between the experimental data for the single-virtual TFF in the
momentum region $Q^2 \gtrsim 2\text{ GeV}^2$ and our prediction
based on the extrapolation of the lattice data using the modified
$z$-expansion is indeed significant. To this end we perform a
combined fit of the same $z$-expansion as used for our global fit in
Sec.~\ref{sec:results}, i.e., using the coefficient set $c_{nm} = \{ c_{00},
c_{10}, c_{11}, c_{22} \}$  with the corresponding lattice-artefact
corrections $\delta_{nm}=\{\delta_{00},\delta_{10}\}$ to the lattice
data, while including all the available experimental data
up to $Q^2 \simeq 35\text{ GeV}^2$. In the $\chi^2/\text{d.o.f.}$ we use equal weights for each set of
experimental data and each set of per-ensemble lattice data. As before,
we perform a model average over all analysis chains and obtain the
AIC-averaged $z$-expansion coefficients
\begin{center}
\begin{tabular}{||c c c c||} 
		\hline
		$c_{00}$ & $c_{10}$ & $c_{11}$ & $c_{22}$ \\
		\hline
		0.2277(31) & -0.0621(52) & -0.1087(85) & 0.776(39) \\
		\hline
\end{tabular}
\end{center}
with the correlation matrix given in
\app{appendix:correlation_latt+exp}. The resulting TFFs for
          diagonal and single-virtual kinematics are shown in
          \Fig~\ref{fig:comb_latt+exp_fit_TFFs}. 
We notice the stability and slightly better determination of the $z$-expansion coefficients $c_{00},
c_{10}$ and the significant shifts in $c_{11}, c_{22}$ with
substantially reduced errors. This leads to a much more precise
determination of the single-virtual TFF at large values of $Q^2
\gtrsim 1.5\text{ GeV}^2$ which is now fully compatible with the
experimental data.
The AIC-averaged value of
$\chi^2/\mathrm{d.o.f.}=1.35$ demonstrates that the single-virtual data from the lattice and
from experiment are indeed completely consistent with each other and
can be described by the same TFF function. For the double-virtual TFF
we note the significant difference for $Q^2 \gtrsim 3.0\text{ GeV}^2$
which is not surprising given the fact that there is no data support
in that momentum range. Additional lattice data is needed there in
order to further restrict the TFF.

Secondly, we can test the stability of the lattice-driven
determination of $\amupi$, $\Gamma(\pi \to \gamma \gamma)$ and
$b_\pi$. For these quantities we obtain from the combined fit to the
lattice and experimental data
\begin{equation}
\begin{gathered}
\amupi = 61.7(2.0)_{\mathrm{stat}}(0.5)_{\mathrm{sys}}[2.0]_{\mathrm{tot}} \times 10^{-11}, \\
\Gamma(\pi \to \gamma \gamma) = 7.97(0.35)_{\mathrm{stat}}(0.11)_{\mathrm{sys}}[0.37]_{\mathrm{tot}}\,\mathrm{eV} ,\\
b_\pi = 2.336(0.049)_{\mathrm{stat}}(0.018)_{\mathrm{sys}}[0.052]_{\mathrm{tot}}\,\mathrm{GeV}^{-2}.
\end{gathered}
\end{equation}
While
these quantities are now much better determined with errors reduced by
factors of 1.4, 1.6 and 3.8, the values are still fully compatible with the
lattice-only determinations in \Eq~(\ref{eq:lattice results}) in
Sec.~\ref{sec:results}, with the largest shift observed in
$\amupi$ upwards by about 1.5$\sigma$. This
is easy to understand from the fact that the experimental data shifts
the TFF to slightly larger values for $Q^2 \gtrsim 2.5\text{ GeV}^2$ outside the
lattice support, and the fact that this momentum region may still
contribute up to 3\% to the total value of $\amupi$, cf., for example,
\Fig\ref{fig:3d_int_rep_saturation}. Nevertheless, this exercise shows
that the low-energy quantities $\amupi$, $\Gamma(\pi \to \gamma \gamma)$ and
$b_\pi$ are not very sensitive to the large-$Q^2$ behavior of the TFF
and can safely be determined from the lattice alone. 

Thirdly, for phenomenological analyses where the large-$Q^2$ behavior
of the TFF is important, the lattice- and data-driven parameterization
given here is most useful. However, we would like to add the
cautionary remark that our analysis does not contain a full systematic
analysis of the dependence on higher-orders in the $z$-expansion
fits. We likely expect that the corresponding systematic error leads
to an increased total error for the TFFs, in particular for those  close to diagonal virtuality where there is no support from lattice or experimental data.

\section{Conclusions and outlook}
\label{sec:conclusion}
We have presented an ab-initio computation of the pion transition form
factor at the physical point in the continuum limit using twisted-mass
lattice QCD, covering the kinematic range relevant for the extraction
of the pion-pole contribution to the HLbL. We are able to include all
disconnected Wick contractions contributing to the amplitudes and
hence relevant for the calculation of the form factors.

This allows us to provide a parameterization of the transition form
factor through the modified $z$-expansion including the determination
of correlations between all coefficients. Our results in the
double-virtual and low-$Q^2$, single-virtual regimes are complementary
to the currently available single-virtual experimental data. Our data
in the latter regime is of particular interest as new experimental
data will become available \cite{Ablikim_2020} which can be compared
with our prediction from first principles.

In the single-virtual regime we find agreement between our calculated
pion form factor and the experimental data up to $Q_2^2 \lesssim
2\text{ GeV}^2$, while the extrapolation to $Q_2^2 \gtrsim 2\text{
  GeV}^2$, based purely on our lattice data, systematically
undershoots the experimental data from CELLO \cite{CELLO:1990klc},
CLEO \cite{CLEO:1997fho}, BaBar \cite{BaBar:2009rrj, BaBar:2011nrp}
and Belle \cite{Belle:Uehara2012},
cf.~\Fig~\ref{fig:comb_fit_TFFs}. In contrast, our results for the
partial decay width $\Gamma(\pi \to \gamma \gamma)$ and the slope
parameter $b_\pi$ are compatible with the experimental values,
cf.~\Fig~\ref{fig:results}. We note, however, that we do not observe
any significant tension between the lattice data and the experimental
one, as discussed in Sec.~\ref{sec:lattice+exp analysis}.

The main result of this paper, namely the pion-pole contribution to HLbL,
\begin{equation}
\amupi = 56.7(3.1)_{\mathrm{stat}}(1.0)_{\mathrm{sys}}[3.2]_{\mathrm{tot}} \times 10^{-11},
\end{equation}
is compatible with recent lattice and data-driven results, with a relative total error at thesub-6\% level. 
The main systematic effect not fully accounted for in this work is the
effect of working at a fixed finite spatial volume. On general grounds
finite-size effects can be estimated in QCD to be of the order
${\cal O}(\exp(-m_\pi L))$, i.e., they are exponentially
suppressed. With our values of $m_\pi L =3.6$--$3.9$ we  expect them to
be at most a few percent. In practice, at the physical lattice sizes of $5.1$--$5.5$\,fm employed in this work, finite-size
effects appear to be negligible compared to the current precision for $\amupi$ and $\Gamma(\pi \to \gamma \gamma)$, 
as demonstrated in Ref.~\cite{Gerardin:2023naa} for similar physical lattice sizes. 
Since there are no lattice results available from previous work for
the slope parameter and its sensitivity to finite-size effects, we can
not exclude it to be affected more. 
We leave the study of finite-size effects as future work, potentially using ETMC ensembles at physical pion mass and the same range of lattice spacings with larger lattice sizes up to $\approx 7.5 \,$fm. 

To achieve further precision in future work, several options are available: One can use momentum $\vec{p} \neq 0$ for the pseudoscalar creation operator (i.e., the moving frame) to give a better coverage of the single-virtual axis. In addition, a fourth physical point ensemble with approximately the same volume as the three used here, but with an even smaller lattice spacing, is currently in production and could then also be included in this calculation.

\begin{acknowledgments}
  We would like to thank all members of the Extended Twisted Mass
  Collaboration (ETMC) for the very enjoyable and fruitful collaboration. 
This work is supported in part by the Sino-German collaborative research center CRC 110 and
the Swiss National Science Foundation (SNSF) through grant
No.~200021\_175761, 200021\_208222, and 200020\_200424. J.F.~received
financial support by the German Research Foundation (DFG) research
unit FOR5269 "Future methods for studying confined gluons in QCD".
J.F., S.B.~and K.H.~received
financial support by the PRACE Sixth Implementation Phase (PRACE-6IP) program (grant agreement
No.~823767) and the EuroHPC-JU project EuroCC (grant agreement No.~951740) of the European Commission. F.P.~acknowledges financial support by the Cyprus Research and Innovation foundation under contracts with numbers EXCELLENCE/0918/0129 and EXCELLENCE/0421/0195.\\
The authors gratefully acknowledge computing time granted on Piz Daint at Centro Svizzero di Calcolo Scientifico (CSCS)
via the projects s849, s982, s1045 and s1133.
The authors also gratefully acknowledge
the Gauss Centre for Supercomputing
e.V. (www.gauss-centre.eu) for funding the project by providing
computing time on the GCS supercomputer JUWELS Booster~\cite{Krause:2019pey} at
the Jülich Supercomputing Centre (JSC) and on the GCS Supercomputer
SuperMUC at Leibniz Supercomputing Centre (www.lrz.de) through the
project pr74yo. 
Part of the results were created within the EA program of JUWELS Booster
also with the help of the JUWELS Booster Project Team (JSC, Atos,
ParTec, NVIDIA). The authors also acknowledge the Texas Advanced Computing Center (TACC) at The University of Texas at Austin for providing HPC resources that have contributed to the research results.
Ensemble production and measurements for this analysis made use of tmLQCD~\cite{Jansen:2009xp,Deuzeman:2013xaa,Abdel-Rehim:2013wba,Kostrzewa:2022hsv}, DD-$\alpha$AMG~\cite{Alexandrou:2016izb,Alexandrou:2018wiv}, and QUDA~\cite{Clark:2009wm,Babich:2011np,Clark:2016rdz}.
Some figures were produced using \texttt{matplotlib}~\cite{Hunter:2007} and \texttt{corner.py}~\cite{corner}.
\end{acknowledgments}

\appendix
\appendix

\section{3d Integral representation weights}
\label{appendix:3d_int_rep}

The weight functions $w_{1,2}$ appearing in Eqs.~\eqref{eq:int_rep_1} and \eqref{eq:int_rep_2} presented here are taken from \cite{PhysRevD.94.053006} and have been derived in \cite{Jegerlehner2009} using the method of Gegenbauer polynomials \cite{PhysRev.136.B1111,PhysRev.163.1699,ROSNER196711,PhysRevD.9.421,PhysRevD.20.2068}.

The weight functions 
read
\begin{align}
w_1(Q_1, Q_2, \tau) &= \left(-\frac{2 \pi}{3} \right) \sqrt{1-\tau^2} \frac{Q_1^3Q_2^3}{Q_2^2+m_\pi^2} I_1(Q_1,Q_2,\tau), \\
w_2(Q_1, Q_2, \tau) &= \left(-\frac{2 \pi}{3} \right) \sqrt{1-\tau^2} \frac{Q_1^3Q_2^3}{(Q_1+Q_2)^2+m_\pi^2} I_2(Q_1,Q_2,\tau),
\end{align}
with
\begin{footnotesize}
\begin{align}
I_1(Q_1,Q_2,\tau) &= X(Q_1,Q_2,\tau)[
		8 P_1 P_2 Q_1 Q_2 \tau - 2 P_1 P_3 (Q_2^4/m_\mu^2 - 2 Q_2^2)  \nonumber \\
		& \quad + 4 P_2 P_3 Q_1^2 - 4 P_2 - 2 P_1 (2 - Q_2^2/m_\mu^2 + 2 Q_1 Q_2 \tau/m_\mu^2) \nonumber \\
		& \quad - 2 P_3 (4 + Q_1^2/m_\mu^2 -2 Q_2^2/m_\mu^2) + 2/m_\mu^2] \nonumber \\
		& - 2 P_1 P_2 (1 + (1 - R_{m1})Q_1 Q_2 \tau/m_\mu^2) \nonumber \\
		& + P_1 P_3 (2 - (1 - R_{m1})Q_2^2/m_\mu^2) \nonumber \\
		& + P_2 P_3 (2 + (1 - R_{m1})^2Q_1 Q_2 \tau/m_\mu^2) \nonumber \\
		& + P_1 (1 - R_{m1})/m_\mu^2 + 3 P_3 (1 - R_{m1})/m_\mu^2
\end{align}
\end{footnotesize}
and
\begin{footnotesize}
\begin{align}
I_2(Q_1,Q_2,\tau) &= X(Q_1,Q_2,\tau)[
		4 P_1 P_2 Q_1 Q_2 \tau + 2 P_1 P_3 Q_2^2 - 2 P_1 \nonumber \\
		& \quad + 2 P_2 P_3 Q_1^2  - 2 P_2 - 4 P_3 - 4/m_\mu^2] \nonumber \\
		& - 2 P_1 P_2 - 3 P_1 (1 - R_{m2})/(2m_\mu^2) - 3 P_2 (1 - R_{m1})/(2m_\mu^2) \nonumber \\
		& - P_3 (2 - R_{m1} - R_{m2})/(2m_\mu^2) \nonumber \\
		& + P_1 P_3 [2 + 3 (1 - R_{m2})Q_2^2/(2m_\mu^2) + (1-R_{m2})^2Q_1 Q_2 \tau/(2m_\mu^2)] \nonumber \\
		& + P_2 P_3 [2 + 3 (1 - R_{m1})Q_1^2/(2m_\mu^2) + (1-R_{m1})^2Q_1 Q_2 \tau/(2m_\mu^2)],
\end{align}
\end{footnotesize}
where
\begin{align}
Q_3^2 &= (Q_1 + Q_2)^2 = Q_1^2 + 2 Q_1 Q_2 \tau + Q_2^2, \\
\tau &= \cos \theta,\\
P_i &= 1/Q_i^2, i \in \{1,2,3\}.
\end{align}
Further,
\begin{align}
X(Q_1Q_2,\tau) &= \frac{1}{Q_1Q_2x} \arctan \left( \frac{zx}{1-z\tau} \right), \\
x &= \sqrt{1 - \tau^2}, \\
z &= \frac{Q_1Q_2}{4 m_\mu^2} (1 - R_{m1})(1 - R_{m2}), \\
R_{mi} &= \sqrt{1 + \frac{4 m_\mu^2}{Q_i^2}}, i \in \{1,2\}.
\end{align}
For a detailed discussion of the behavior of $w_{1,2}$ in different
limits see \cite{PhysRevD.94.053006}.

\section{Correlation matrix for the combined fit to lattice and
  experimental data}
\label{appendix:correlation_latt+exp}
Here we provide the correlation matrix for the combined fit of the
modified $z$-expansion as desrcibed in Sec.~\ref{sec:lattice+exp
  analysis} using the coefficient set  $c_{nm} = \{ c_{00},
c_{10}, c_{11}, c_{22} \}$  with the corresponding lattice-artefact
corrections $\delta_{nm}=\{\delta_{00},\delta_{10}\}$ to the lattice
and experimental data,
\begin{equation}
\label{eq:cont_corrmat}
\mathrm{cor}(c_{nm}) = \left(
  \begin{array}{cccc} 
    +1.00 & -0.14 & +0.54 & -0.78 \\
    -0.14 & +1.00 & -0.12 & -0.06 \\
    +0.54 & -0.12 & +1.00 & +0.07 \\
    -0.78 & -0.06 & +0.07 & +1.00 
  \end{array} \right) \, .
\end{equation}
We refrain from providing the corresponding corner plot, since it
serves the purpose of illustration only.

\bibliographystyle{JHEP}
\bibliography{bibliography}{}
\end{document}